\documentclass[10pt,conference]{IEEEtran} 

\usepackage{tikz}
\usepackage{amsmath}

\usepackage{filecontents}
\usepackage{booktabs}
\usepackage{amsthm}

\definecolor{cblue}{rgb}{0.0, 0.28, 0.67}

\usepackage{subfigure}
\usepackage{url}
\usepackage{mathtools}
\usepackage{amsfonts}
\usepackage{amssymb}
\usepackage{multirow}
\usepackage{algorithm}
\usepackage{algpseudocode}
\usepackage{graphicx}
\usepackage{enumitem}
\usepackage[bookmarks=false]{hyperref}
\hypersetup{draft}
\usepackage{balance}

\usepackage{caption}  

\newtheorem{definition}{Definition}

\usepackage{lmodern}
\DeclareMathDelimiter{(}{\mathopen} {operators}{"28}{largesymbols}{"00}
\DeclareMathDelimiter{)}{\mathclose}{operators}{"29}{largesymbols}{"01}

\setitemize{noitemsep,topsep=0pt,parsep=0pt,partopsep=0pt}
\setdescription{noitemsep,topsep=0pt,parsep=0pt,partopsep=0pt}

\makeatletter
\g@addto@macro{\normalsize}{%
    \setlength{\abovedisplayskip}{5pt}
    \setlength{\abovedisplayshortskip}{5pt}
    \setlength{\belowdisplayskip}{5pt}
    \setlength{\belowdisplayshortskip}{5pt}}
\makeatother

\usepackage{lmodern}
\newcommand{\Addliquidity}{{\color{blue}\normalfont\texttt{AddLiquidity}}}
\newcommand{\addliquidity}{{\color{blue}\normalfont\texttt{addLiquidity}}}

\newcommand{\Removeliquidity}{{\color{blue}\normalfont\texttt{RemoveLiquidity}}}
\newcommand{\removeliquidity}{{\color{blue}\normalfont\texttt{removeLiquidity}}}

\newcommand{\TransactXY}{{\color{blue}\normalfont\texttt{Transact$X$for$Y$}}}
\newcommand{\transactXY}{{\color{blue}\normalfont\texttt{transact$X$for$Y$}}}
\newcommand{\TransactYX}{{\color{blue}\normalfont\texttt{Transact$Y$for$X$}}}

\DeclareMathOperator{\E}{\mathbb{E}}

\usepackage{pdflscape}
\usepackage{rotating}
\usepackage{array}
\newcolumntype{L}{>{\centering\arraybackslash}m{3cm}}

\usepackage{pifont}
\newcommand{\cmark}{\ding{51}}%
\newcommand{\xmark}{\ding{55}}%

\usepackage{lipsum}

\newcommand\blfootnote[1]{%
  \begingroup
  \renewcommand\thefootnote{}\footnote{#1}%
  \addtocounter{footnote}{-1}%
  \endgroup
}

\begin{document}

\title{High-Frequency Trading on \\Decentralized On-Chain Exchanges}

\author{
\IEEEauthorblockN{
Liyi Zhou \IEEEauthorrefmark{1}, 
Kaihua Qin \IEEEauthorrefmark{1}, 
Christof Ferreira Torres \IEEEauthorrefmark{2}, 
Duc V Le \IEEEauthorrefmark{3} 
and 
Arthur Gervais \IEEEauthorrefmark{1}
}

\IEEEauthorblockA{
\IEEEauthorrefmark{1}Imperial College London, United Kingdom\\
}
\IEEEauthorblockA{
\IEEEauthorrefmark{2}University of Luxembourg, Luxembourg\\
}
\IEEEauthorblockA{
\IEEEauthorrefmark{3}Purdue University, United States\\
}
}

\maketitle
\thispagestyle{plain}
\pagestyle{plain}

\begin{abstract}
Decentralized exchanges (DEXs) allow parties to participate in financial markets while retaining full custody of their funds. However, the transparency of blockchain-based DEX in combination with the latency for transactions to be processed, makes market-manipulation feasible. For instance, adversaries could perform \textit{front-running}~---~the practice of exploiting (typically non-public) information that may change the price of an asset for financial gain.

In this work we formalize, analytically exposit and empirically evaluate an augmented variant of front-running: \textit{sandwich attacks}, which involve front- and back-running victim transactions on a blockchain-based DEX. We quantify the probability of an adversarial trader being able to undertake the attack, based on the relative positioning of a transaction within a blockchain block. We find that a single adversarial trader can earn a daily revenue of over several thousand USD when performing sandwich attacks on one particular DEX~---~Uniswap, an exchange with over $5$M USD daily trading volume by June 2020. In addition to a single-adversary game, we simulate the outcome of sandwich attacks under multiple competing adversaries, to account for the real-world trading environment.
\end{abstract}

\section{Introduction}\label{sec:introduction}
\blfootnote{Disclosure: Arthur Gervais works on the Liquidity Network, a community-driven, open source layer-2 blockchain scaling solution.}
Decades of asset trading on traditional exchanges have brought to fruition a veritable collection of market manipulation techniques, such as front-running~\cite{front-running}, pump and dump schemes~\cite{xu2019anatomy} and wash trading~\cite{blockchaintransparency}. In the context of cryptocurrencies, research to date indicates that the ecosystem requires a greater awareness of such malpractices~\cite{xu2019anatomy, daian2019flash, mavroudis2019libra}, and better exchange design~\cite{bentov2019tesseract} to prevent misbehavior. Most existing legislation does not regulate crypto-exchanges to the same degree as traditional exchanges~---~leaving ignorant traders open to exploitation by predatory practices, some of which is close to risk-free.

Decentralized exchanges (DEXs) allow traders to trade financial assets without giving up asset custody to a third party. Orders can be placed and matched in their entirety through immutable blockchain smart contracts, offering the possibility of censorship resistance, where orders cannot be modified prior and after execution\footnote{DEX prevent anyone from censoring trades, even the exchange itself. Censorship-resistance is a key property of permissionless blockchains such as Bitcoin and Ethereum, where users remain custodian over their assets at any point in time.}. Censorship-resistant trade is itself made possible through reliance on an underlying blockchain, which makes public all attempted and executed trades within its peer-to-peer (P2P) network. The transparency of the blockchain layer, however, in combination with the latency for orders to deterministically execute makes, front-running easier to undertake~---~and hence influences negatively the security of the trader's assets.

\textbf{This paper.}
We focus on a combination of front- and back-running\footnote{While the SEC defines front-running as an action on \emph{private} information, we only operate on \emph{public} trade information.}, known as a \textit{sandwiching}, for a single on-chain DEX. To the best of our knowledge, we are the first to formalize and quantify sandwich attacks. To make their sandwich, a predatory trader first observes a blockchain P2P network for a victim transaction and then rushes to squeeze it by placing one order just before the transaction (i.e.\ front-run) and one order just after it (i.e.\ back-run). If the target transaction is going to increase (decrease) the price of an asset, the adversary can place an order before which buys (sells) the asset in question, and an order afterward which sells (buys) the asset again.

We restrict our focus to \emph{automated market maker} (AMM) DEXs~\cite{uniswap2018, bancor-front-running}, as opposed to DEXs which operate limit order books (LOB)~\cite{Commission}, on account of their deterministic nature which enables us to rely on fewer assumptions in our analysis. AMM DEXs simplify trading by algorithmically performing market making\footnote{The process of serving a market with the possibility to purchase and sell an asset.}, resulting in near-instant liquidity (i.e.\ the ability to purchase and sell assets) for market participants. Uniswap is a prominent example of an AMM DEX, which, by March 2020, has amassed a total liquidity of nearly $48$M USD (corresponding to a 75\% market liquidity share for AMM DEX) and had a trading volume of over $250$M USD since its inception in November $2018$. We formalize, analytically exposit and empirically evaluate \textit{sandwiching} on AMM DEXs. We quantify optimal adversarial revenues and perform a real-world empirical evaluation of sandwich attacks. We also study the probability of a transaction having a particular relative position within a blockchain block, informing the prospects for such an attack. Finally, to account for a real-world scenario in which multiple adversaries are likely to compete over victim transactions, we perform simulations to quantify the transaction fees resulting from a reactive fee counter-bidding contest.

\noindent \textbf{Summary of contributions:} 
\begin{itemize}
    \item \textbf{Formalization of sandwich attacks.}
    We state a mathematical formalization of the AMM mechanism and the sandwich attack, providing an adversary with a framework to manage their portfolio of assets and maximize the profitability of the attack.
    \item \textbf{Analytic and empirical evaluation.}
    We analytically and empirically evaluate sandwich attacks on AMM DEX. Besides an adversarial liquidity taker, we introduce a new class of sandwich attacks performed by an adversarial liquidity provider. We quantify the optimal adversarial revenue and validate our results on the Uniswap exchange (largest DEX, with $5$M USD trading volume at the time of writing). Our empirical results show that an adversary can achieve an average daily revenue of $3,414$ USD\footnote{We disclosed our preliminary results to Uniswap on $18$th of November $2019$, which allowed tightening the trader protections.}.
    Even without collusion with a miner, we find that, in the absence of other adversaries, the likelihood to position a transaction \emph{before} or \emph{after} another transaction within a blockchain block is at least $79\%$, using a transaction fee payment strategy of $\pm 1$ Wei\footnote{The smallest amount of Ether currency, $1$ WEI $=10^{-18}$ ETH}.
    \item \textbf{Multiple Attacker Game.}
    We simulate the sandwich attacks under multiple simultaneous attackers that follow a reactive counter-bidding strategy~\cite{daian2019flash}. We find that the presence of $2$, $5$ and $10$ attackers respectively reduce the expected profitability for each attacker by $51.0$\%, $81.4$\% and $91.5$\% to $0.45$, $0.17$, $0.08$ ETH ($67$, $25$, $12$ USD), given a victim that transacts $20$ ETH to DAI on Uniswap with a transaction pending on the P2P layer for $10$ seconds before being mined. If the blockchain is congested (i.e.\ the victim transaction remains pending for longer than the average block interval), we show that the break-even of the attacker becomes harder to attain.
    \item \textbf{DEX security vs.\ scalability tradeoff.}
    Our work uncovers an inherent tension between the security and scalability of an AMM DEX. If the DEX is used securely (i.e.\ under a low or zero price slippage), trades are likely to fail under high transaction volume; and an adversarial trader may profit otherwise.
\end{itemize}


\section{Decentralized Exchanges}\label{sec:background}
At the root of decentralized exchanges are blockchains. Blockchains, such as Bitcoin~\cite{bitcoin}, enable peers to transact without trusting third-party intermediaries. The core component of a blockchain is a hash-linked chain of blocks~\cite{bonneau2015sok}, where miners form blocks as a data-structure which accumulates transactions. Blockchains which allow the execution of smart contracts~\cite{wood2014ethereum}, constitutes the basic building block for exchanges. A crucial aspect of this paper is that in most blockchain designs, transactions are executed in the sequence in which they are written into a block. This sequence dependence matters for blockchain-based exchanges, and will be detailed extensively in Section~\ref{sec:sandwich-attack}. An exchange is built out of three main components: a price discovery mechanism, a trade matching engine, and a trade clearing system. Blockchains allow these components to be encoded within a smart contract to construct a decentralized, or non-custodial exchange~\cite{hertzog2017bancor, uniswap2018, kyber2019, oasis2019, idex2019, dutchx2019}. The non-custodial property guarantees that a trader retains custody over their assets at any point in time. If all exchange components are implemented within smart contracts, the exchange qualifies as an \emph{on-chain DEX}. If only the trade clearing is realized within a smart contract, the exchange may be centralized but can retain the non-custodial property~\cite{idex2019}.

\subsection{DEX components}\label{sec:dex-components}
A DEX is a game between a liquidity provider and taker.

\begin{description}
    \item[Liquidity Provider:] a market participant that provides liquidity (financial asset trade offers).
    \item[Liquidity Taker:] a market participant that buys or sells one asset in exchange for another asset, by taking the liquidity offered by a liquidity provider.
\end{description}

Further, we distinguish between two varieties of DEX, depending on their mechanisms of price discovery.

\begin{description}
\item[Order Book:] a list of buy and sell orders for a particular asset, where each order stipulates a price and and quantity. A liquidity provider quotes bid and ask prices, with an associated volume, competing for liquidity taker order flow~\cite{kyber2019, oasis2019, idex2019}, such that a match between supply (from a liquidity provider) and demand (from a liquidity taker) is facilitated (also referred to as \emph{market making}).
\item[Automated Market Maker (AMM):] A predefined pricing algorithm automatically performs price-discovery and market making, using assets within liquidity pools~\cite{uniswap2018, bancor-front-running}. Liquidity providers are, therefore, not required to monitor the market to adjust bid and ask prices. Liquidity takers can directly trade against the AMM liquidity. Such automation also serves to reduce the number of on-chain transactions, making such mechanisms particularly suitable for smart contract-based DEXs given an underlying blockchain that supports only a limited number of transactions per second (tps).
\end{description}

\subsection{AMM Mechanism}\label{sec:amm-mechanism}
We denote with $X$/$Y$ an asset pair composed of asset $X$ and $Y$. An AMM asset pair $X$/$Y$ consists of two liquidity pools, respectively for each asset:

\begin{description}
\item[Asset $X$ liquidity pool ($x \in \mathbb{R}^{+}$):] Total amount of asset $X$ deposited by liquidity providers.
\item[Asset $Y$ liquidity pool ($y \in \mathbb{R}^{+}$):] Total amount of asset $Y$ deposited by liquidity providers.
\end{description}

\begin{definition}\label{def:amm-state}
The state (or depth) of an AMM market $X$/$Y$ is defined as $(x, y)$, $x$ the amount of asset $X$, $y$ the amount of asset $Y$ in the liquidity pool. The state at a given blockchain block $N$ is denoted $(x_N, y_N)$.
\end{definition}

AMM DEXs support the following actions.

\begin{description}
    \item[\Addliquidity:] A liquidity provider deposits $\delta_x$ of asset $X$, and $\delta_y$ of asset $Y$ into the corresponding liquidity pools (cf.\ Equation~\ref{eq:AMM_add_liquidity}).
    \begin{equation}\label{eq:AMM_add_liquidity}
     (x,y) \xrightarrow[\delta_x \in \mathbb{R^{+}},\ \delta_y \in \mathbb{R^{+}}]{\Addliquidity(\delta_x, \delta_y)} (x + \delta_x, y + \delta_y)   
    \end{equation}
    \item[\Removeliquidity:] A liquidity provider withdraws $\delta_x$ of asset $X$, and $\delta_y$ of asset $Y$ from the corresponding liquidity pools (cf.\ Equation~\ref{eq:AMM_remove_liquidity}).
    \begin{equation}\label{eq:AMM_remove_liquidity}
    (x,y) \xrightarrow[\delta_x \in \mathbb{R}^{+} \leq x,\ \delta_y \in \mathbb{R}^{+} \leq y]{\Removeliquidity(\delta_x, \delta_y)} (x - \delta_x, y - \delta_y)
    \end{equation}
    \item[\TransactXY:] A liquidity taker can trade $\delta_x$ of asset $X$, increasing the available liquidity of asset $X$, in exchange for $\delta_y = f(\delta_x - c_x(\cdot)) - c_y(\cdot)$ of asset $Y$, decreasing the available liquidity of asset $Y$ (cf.\ Equation~\ref{eq:AMM_transact_y_for_x}). $c_x(\cdot), c_y(\cdot)$ represent the trade fees in asset $X$ and $Y$ respectively. $f(\cdot)$ calculates the amount of asset $Y$ purchased by the liquidity taker. Each AMM exchange may chose a custom pricing function $f(\cdot)$ for governing the asset exchange~\cite{balancerexchange}. Note that the exchange asset pricing cannot be determined by a simple constant, as the market dynamics of purchasing and selling power must be modeled within the exchange (i.e.\ the more assets on would want to purchase, the higher the fees).
    \begin{equation}\label{eq:AMM_transact_y_for_x}
    (x,y) \xrightarrow[\delta_x \in \mathbb{R}^{+}]{\TransactXY(\delta_x)} (x + \delta_x,\ y - f(\delta_x - c_x(\cdot)) + c_y(\cdot))
    \end{equation}
\end{description}

\textbf{Constant Product AMM.}
The simplest AMM mechanism is a constant product market maker, which keeps the product $x \times y$ constant for any arbitrary asset pair ($X/Y$). In this work we focus on the constant product model, because with over 75\% market liquidity, this represents the most prevalent AMM model. In the following, $k$ denotes the product of supplies ($k \in \mathbb{R}^{+} = x \times y$), which remains constant upon taker transactions. $k$ changes when a liquidity provider deposits, or withdraws $X$/$Y$ pool funds. Equation~\ref{eq:Constant_AMM_transact_x_for_y} shows the state changes of \TransactXY\ under a constant product AMM.

\begin{equation}\label{eq:Constant_AMM_transact_x_for_y}
(x,y) \xrightarrow[\delta_x \in \mathbb{R}^{+}]{\TransactXY(\delta_x)} (x + \delta_x,\ \frac{xy}{x + \delta_x - c_x(\cdot)} + c_y(\cdot))
\end{equation}

\subsection{Price Slippage}\label{sec:price-slippage}
\begin{figure}[htb!]
\centering
\includegraphics[width = \linewidth]{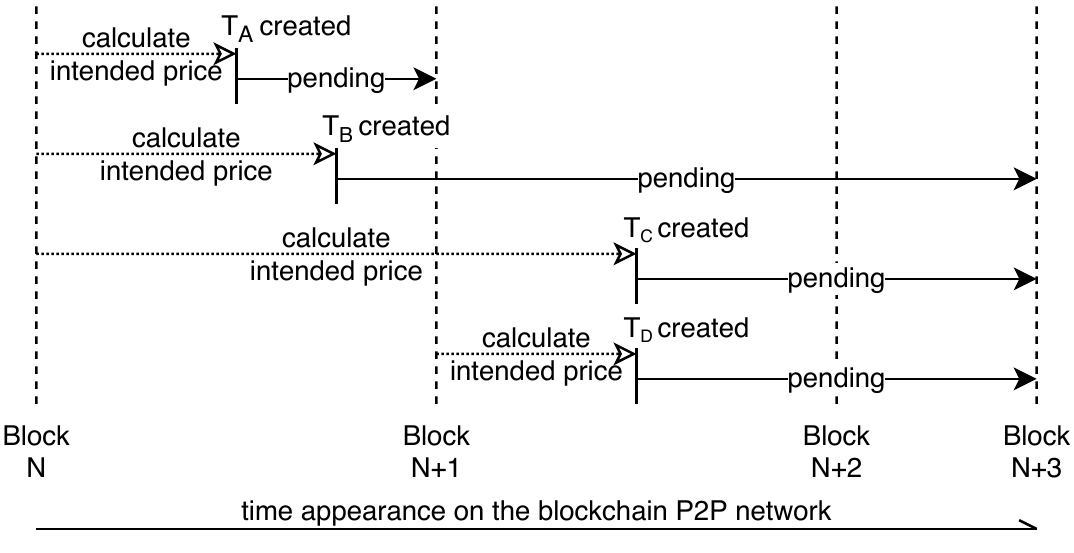}
\caption{Visualizing the cause of unexpected slippage. $\E{[P]}$ of $T_A$ is based on the AMM state of block $N$. $T_A$ does not suffer from unexpected slippage, because no concurrent transactions exist. $T_B$ executes in block $N+3$. $\E{[P]}$ of $T_C$'s is based on block $N$, as we assume network delays. If $T_C$ and $T_D$ change the state of the underlying market, those may induce unexpected slippage for $T_B$.}
\label{fig:AMM_slippage}
\end{figure}

Price slippage is the change in the price of an asset during a trade. \emph{Expected price slippage} is the expected increase or decrease in price based on the volume to be traded and the available liquidity~\cite{investopedia-slippage}, where the expectation is formed at the beginning of the trade. The higher the quantity to be traded, the greater the expected slippage (cf.\ Table~\ref{tab:slippage-example}). \emph{Unexpected price slippage} refers to any \emph{additional} increase or decrease in price, over and above the expected slippage, during the interveni period from the submission of a trade commitment to its execution. This can be thought of as an expectation error. When an exchange's market liquidity changes, the resulting \emph{actual} slippage is challenging to foresee (cf.\ Figure~\ref{fig:AMM_slippage}), making the formation of accurate expectations more challenging. We note the following definitions.  

\begin{description}
    \item[Expected Execution Price ($\E{[P]}$):] When a liquidity taker issues a trade on $X$/$Y$, the taker wishes to execute the trade with the expected execution price $\E{[P]}$ (based on the AMM algorithm and $X$/$Y$ state), given the expected slippage.
    \item[Execution Price ($P$):] During the time difference between a liquidity taker issuing a transaction, and the transaction being executed (e.g.\ mined in a block), the state of the AMM market $X$/$Y$ may change. This state change may induce unexpected slippage resulting in an execution price $P \neq \E{[P]}$.
    \item[Unexpected Price Slippage ($P - \E{[P]}$):] is the difference between $P$ and $\E{[P]}$.
    \item[Unexpected Slippage Rate ($\frac{P - \E{[P]}}{\E{[P]}}$):] is the unexpected slippage over the expected price.
\end{description}

\begin{table}[b!]
\centering
\resizebox{\columnwidth}{!}{
\setlength\tabcolsep{5pt}
\begin{tabular}{@{}lcccc@{}}
\toprule
                          & \multicolumn{2}{c}{AMM State 1} & \multicolumn{2}{c}{AMM State 2} \\ 
\midrule
Liquidity $X$ ($x$)       & \multicolumn{2}{c}{$100$}         & \multicolumn{2}{c}{$1,000$} \\
Liquidity $Y$ ($y$)       & \multicolumn{2}{c}{$10$}          & \multicolumn{2}{c}{$100$} \\
Product ($k = xy$)        & \multicolumn{2}{c}{$1,000$}       & \multicolumn{2}{c}{$100,000$} \\
Purchase amount X         & $1$ & $10$                          & $1$ & $10$ \\
AMM Price Y/X             & $0.1000$ & $0.1000$                 & $0.1000$ & $0.1000$\\
$\E{[P]}$ Y/X & $0.1010$ & $0.1111$                 & $0.1001$ & $0.1010$\\
Expected slippage         & $0.0010$ & $0.0111$                 & $0.0001$ & $0.0010$\\
Slippage rate             & $1.01\%$ & $11.11\%$                & $0.10\%$ & $1.01\%$\\
\bottomrule
\end{tabular}
}
\caption{Example price slippages on an AMM DEX.}
\label{tab:slippage-example}
\end{table}

\paragraph{Slippage Example}\label{app:slippage-example}
For example, a liquidity taker, who intends to trade $1$ asset $X$ for $20$ $Y$ at an exchange, results in a price of $0.05$, quoted in units of asset $Y$. However, by the time the AMM DEX executes this transaction, if the price increases to $0.1$, the liquidity taker would only receive $10$ $Y$ for $1$ $X$. The unexpected slippage, in this case, is $0.1 - 0.05 = 0.05$. Slippage can also be negative, i.e.\ a liquidity taker can receive more asset $Y$ than expected. If the execution price above decreases to $0.25$, the liquidity taker would receive $40$ $Y$ for $1$ $X$, with a corresponding unexpected slippage of $0.1 - 0.25 = -0.15$.

\section{Sandwich Attacks on AMM DEXs}\label{sec:sandwich-attack}
In traditional financial markets, the predatory trading strategy of \emph{front-running} involves exploiting (typically non-public) information about a pending trade, expected to materially change the price of an asset, by buying or selling the asset beforehand~\cite{front-running}. If the asset is expected to rise (fall) in price as a result of the trade, the front-runner will seek to buy (sell) the asset before the large pending transaction executes. AMM DEXs aim to mitigate malpractice by providing complete transparency about the available liquidity for assets $X$, $Y$, all pending and performed trades, and therefore removing the role played by non-public information. However, AMM DEXs also exacerbate malpractices by quoting asset prices in a fully deterministic way, providing relative certainty over the expected price impact of a trade. This enables a front-running adversary to perform attacks with predictable outcomes. In the following, we study two sandwich attacks on constant product AMM asset exchanges:

\begin{description}
    \item 1) Liquidity taker attacks liquidity taker.
    \item 2) Liquidity provider attacks liquidity taker.
\end{description}

In each case, the fundamental intuition is that the delay in the time taken for a transaction to execute allows an adversary to profit by exploiting the knowledge of the direction of a price change. The attacks are called sandwich attacks because a victim transaction is \emph{sandwiched} between adversarial transactions.

\begin{figure}[htb!]
\begin{center}
\includegraphics[width = \columnwidth]{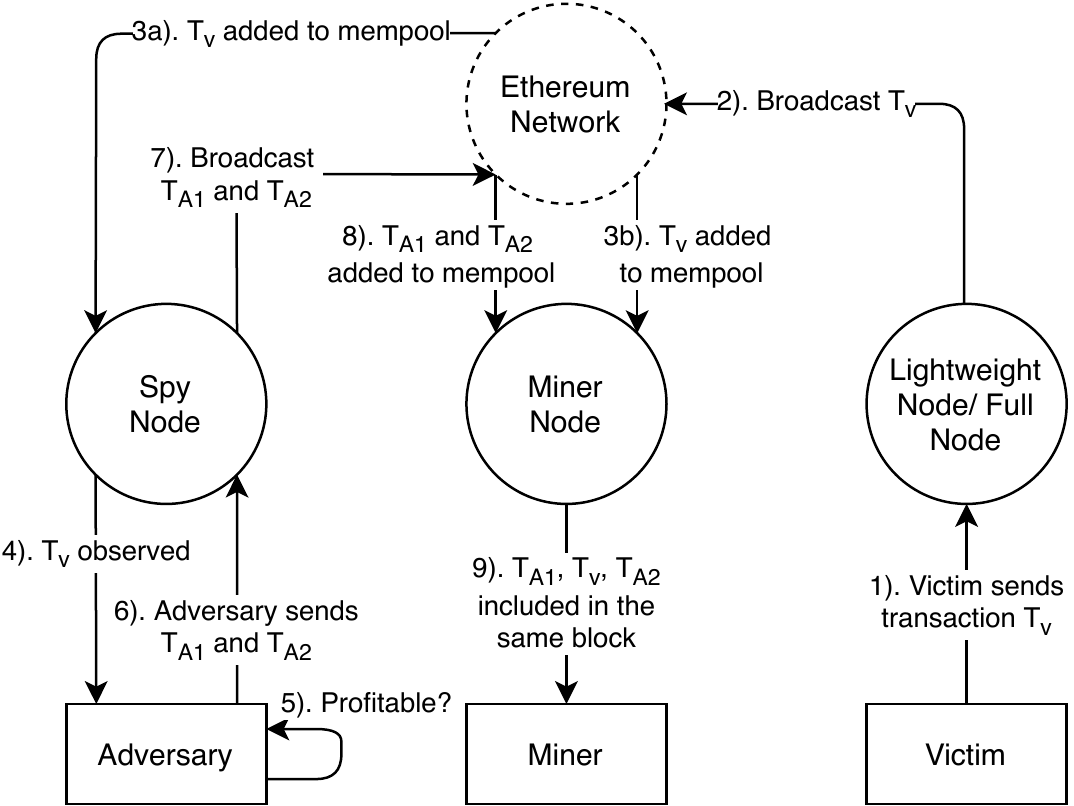}
\end{center}
\caption{Sandwich attack system.}
\label{fig:system}
\end{figure}

\subsection{System Model}
We consider a blockchain P2P network, where a victim initiates trades on an AMM DEX (cf.\ Figure~\ref{fig:system}). An adversary observes pending victim transactions (i.e.\ not yet mined transactions within the memory-, or mempool) through a spy node (e.g.\ a custom Ethereum client), and a miner chooses to include transactions within a block according to a policy (cf.\ Section~\ref{sec:block-position}). A victim transaction trades a crypto-currency asset (such as ETH, DAI, SAI, VERI) to another crypto-asset. We do not consider blockchain forks. While blockchains typically provide delayed finality after $k$ blocks~\cite{garay2015bitcoin, gervais2016security}, we consider a transaction final once included within a block.

\subsection{Threat Model}\label{sec:threat-model}
We consider one computationally bounded and economically rational adversary $A$ (cf.\ Section~\ref{sec:multiple-attacker} for an extended threat model with multiple adversaries), that observes a zero-confirmation transaction $T_V$ from a victim trader $V$ on a blockchain P2P network. The adversarial trader can issue its own transaction $T_{A, f}$ with a transaction fee $f$. Depending on $f$, and the age of propagation, $T_{A, f}$ may be included within the blockchain prior or past $T_V$ (cf.\ Section~\ref{sec:block-position}). In this work we focus on these novel cases where the adversary is not colluding with a miner, i.e.\ we weaken the adversary to quantify a lower bound on the feasibility and profitability of the proposed attacks. Outside of this work, a (stronger) adversary may collude, or bribe a miner~\cite{liao2017incentivizing,mccorry2018smart,bonneau2016buy}, to influence the  transactions ordering within a block, or even to fork the chain as in to discard unsuccessful attacks. We moreover assume that an attack against one victim transaction is independent from other concurrent attacks towards other victim transactions.

\subsection{Liquidity Taker Attacks Taker}\label{sec:sandwich-attack1}
\begin{figure}[htb!]
\centering
\includegraphics[width = 0.95\linewidth]{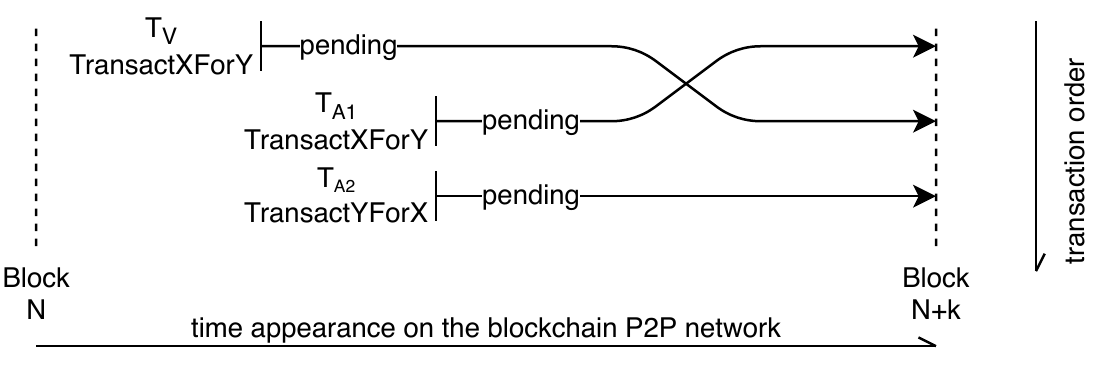}
\caption{An adversarial liquidity taker $A$ attacks a victim taker $V$ on an AMM DEX. Transaction $T_V$ specifies its slippage protection based on the AMM state of block $N$. The adversary's goal is to include $T_{A1}$, $T_V$ and $T_{A2}$ in the same block $N + k, k \in \mathbb{Z}^+$ in that sequence.}
\label{fig:SandwichAttackLTvsLT}
\end{figure}

In our first attack, a liquidity taker targets a victim liquidity taker who has emits on the blockchain P2P network an AMM DEX transaction (\TransactXY), formalizing~\cite{bancor-front-running}. The adversary then emits two transactions (one front- and one back-running) to exploit the victim transaction~$T_V$ (cf.\ Figure~\ref{fig:SandwichAttackLTvsLT}). These three transactions are then unconfirmed in the blockchain P2P network, until a miner choses to include and execute them within a block. The adversary can influence the position of the adversarial transactions, relative to the victim transaction, by paying a higher, or lower transaction fee amount (cf.\ Section~\ref{sec:block-position}).

We refer the interested reader to the Appendix~\ref{app:taker-attacks-taker} for the technical details of the involved transactions.

\subsection{Liquidity Provider Attacks Taker}\label{sec:sandwich-attack2}
\begin{figure}[htb!]
\centering
\includegraphics[width = \linewidth]{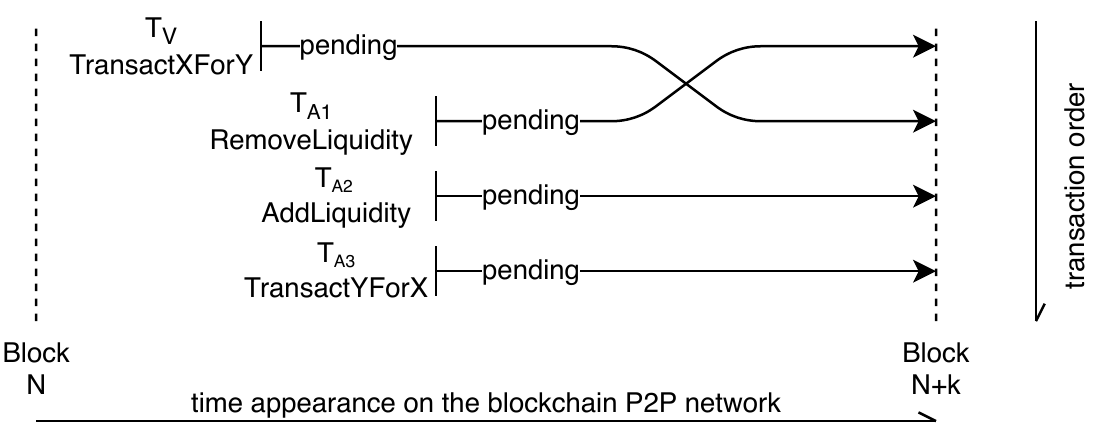}
\caption{An adversarial liquidity provider $A$ attacks a victim taker $V$. $T_V$ transacts asset $Y$ for asset $X$.}
\label{fig:SandwichAttackLPvsLT}
\end{figure}

We present a novel sandwich attack where a liquidity \emph{provider} targets a victim liquidity taker transaction (\TransactXY) on the blockchain P2P network. Upon observing the victim transaction, the adversary emits three transactions (cf.\ Figure~\ref{fig:SandwichAttackLPvsLT}): 
\begin{enumerate}
    \item \Removeliquidity\ (increases victim's slippage)
    \item \Addliquidity\ (restores pool liquidity)
    \item \TransactYX\ (restores asset balance of X)
\end{enumerate}
The \emph{(i)} front-running \Removeliquidity\ transaction reduces the market liquidity of the AMM DEX and increases the victim's unexpected slippage. The \emph{(ii)} back-running \Addliquidity\ transaction restores the percentage of liquidity $A$ holds before the attack. Finally, \emph{(iii)} the back-running transaction \TransactYX\ equilibrates the adversary's balance of asset X to the state before the attack.

Note that liquidity providers earn commission fees proportional to the liquidity (i.e.\ the amount of assets) they provide to an AMM DEX market. In this attack, the adversary $A$ withdraws all its assets from the liquidity pool before $T_V$ executes. As such, $A$ foregoes the commission fees for the victim's transaction. We refer the interested readers to Appendix \ref{app:provider-attacks-taker} for further technical details.

\subsection{Model Limitations}\label{sec:limitation}

\subsubsection{Margin and Leveraged Trading}
Margin trading is the process of using borrowed funds to amplify trading profits (or losses). A trader commits a percentage of the total trade value to open a margin position. For example, to open a $5 \times$ short ETH for DAI position with $10$ETH, the trader needs to commit $2$ETH as collateral. A short position reflects the expectation that the ETH price will decrease, whereas a long position reflects the opposite. The margin platform will then lend to the trader $10$ETH and convert those assets to DAI. If the ETH price decreases, the trader can close the margin trade with a profit.

A limitation of our work is that we do not consider on-chain margin platforms utilizing AMM exchanges to open short/long positions (e.g.\ the recently attacked bZx platform~\cite{qin2020attacking}). An on-chain margin trade system would enable an adversary to reduce the capital requirements for sandwich attacks, at the cost of higher transaction fees (for opening and closing margin trades). Margin trading is unlikely to affect the adversary's monetary revenue because the victim configures a fixed slippage.

\subsubsection{Blockchain Forks}
We do not consider the impact of stale blocks in our analysis. In practice, it is possible that a transaction is included in a stale block (on the forked chain), but is not included in the confirmed blocks (on the main chain). This stale transaction is typically re-injected into the blockchain client's mempool when the stale block is added as an uncle to the main chain. The stale re-injection process of adversarial and victim transaction may increase the failure rate of sandwich attacks, but we leave quantitative results for future work.

\section{Analytical Evaluation}\label{sec:analytical-evaluation}
In this section, we perform the analytical evaluation of sandwich attacks on Uniswap~\cite{uniswap2018}. Uniswap is the most popular DEX at the time of writing with $1,301$ markets, on average $7.30$ provider per market and $29.3$M USD liquidity. From Uniswap's inception in November $2018$ to November $2019$, we identified a trade volume of $1.6$M ETH ($248$M USD), measured on a full archive Geth node ($6$-core Intel i$7$-$8700$ CPU, $3.20$GHz, $64$GB RAM, $10$TB SSD)\footnote{We focus only on transactions executed on Uniswap endpoints directly, not internal transactions that are routed to Uniswap.}. In what follows, we base our evaluations on Uniswap parameters and adopt its liquidity pool distributions~\cite{uniswap2018} from Ethereum block $9$M (mined $25$th November $2019$). In this section we present the analytical results for the two sandwich attacks presented in Section~\ref{sec:sandwich-attack1} and~\ref{sec:sandwich-attack2}.

\begin{figure*}[h!]
\centering
\subfigure[ETH/SAI market]{%
\label{fig:revenue_attack_1_sai}%
\includegraphics[height=2.8in]{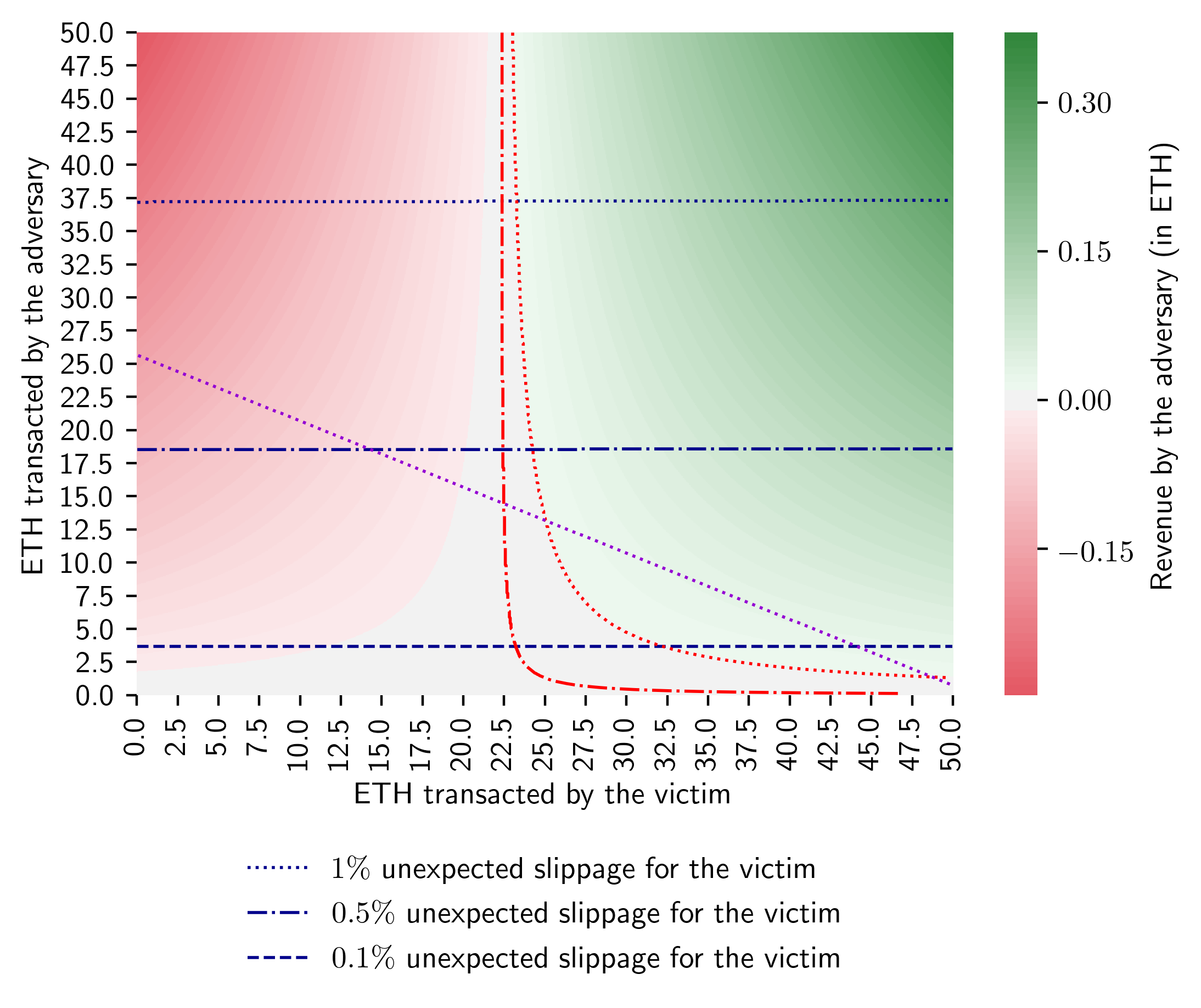}}%
\qquad
\subfigure[ETH/DAI market]{%
\label{fig:revenue_attack_1_dai}%
\includegraphics[height=2.8in]{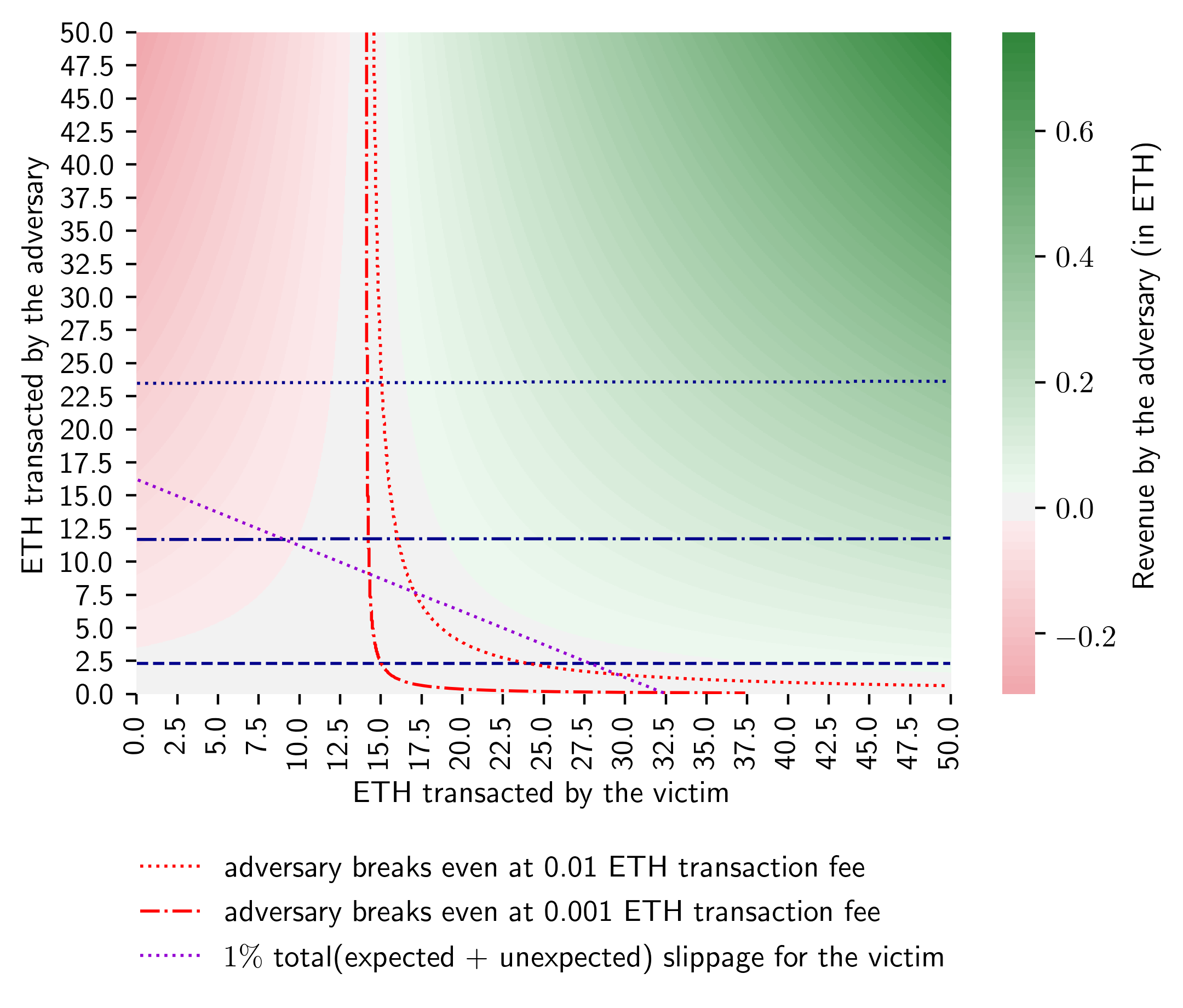}}%
\caption{Analytical sandwich attack by a liquidity taker on a taker (Uniswap, block $9$M, $0.3\%$ fees, $0.5\%$ unexpected slippage). If $T_V$ transacts $40$ ETH for SAI, $A$ gets a max.\ revenue by front-running $T_V$ with a trade $18.59$ ETH for $2,754.32$ SAI, and back-running with $2,754.32$ SAI for $18.68$ ETH. This results in a profit of $0.08$ ETH ($11.74$ USD), if $A$ bears $0.01$ ETH tx fees. Note that the two sub-legends of each figure apply to both sub-figure~\ref{fig:revenue_attack_1_sai} and~\ref{fig:revenue_attack_1_dai}.}
\end{figure*}

\subsection{Adversarial Liquidity Taker} \label{sec:adversarial_liquidity_taker}
At Ethereum mainnet block $9$M, the ETH/SAI Uniswap offers $7,377.53$ ETH and $521,468.62$ SAI\footnote{In the following we adopt the ETH/SAI exchange rate at block $9$M as the ETH/USD exchange rate. $1$ ETH $= 148.97$ USD}. The ETH/DAI Uniswap offers $4,660.75$ ETH and $693,706.47$ DAI. Given this market information, and the constant product formula (cf.\ Section~\ref{sec:background}), we plot in Figure~\ref{fig:revenue_attack_1_sai} and~\ref{fig:revenue_attack_1_dai} the revenue of an adversarial taker performing a sandwich attack against another taker. We visualize three unexpected slippage thresholds ($0.1\%$, $0.5\%$ and $1\%$). We plot the lines at which an adversary would break even given a total ($T_{A1}$ and $T_{A2}$) of $0.01$ ETH ($1.97$ USD) and $0.001$ ETH ($0.2$ USD) worth of transaction fees\footnote{At the time of writing, the average Ethereum transaction fee is $0.13$ USD (\url{https://bitinfocharts.com/comparison/ethereum-transactionfees.html})}. We observe that the greater the amount of ETH transacted by the victim, the greater is the adversarial revenue. For example, given an unexpected slippage protection of $0.5\%$, an adversarial taker gains a revenue of $0.01$ ETH ($2.03$ USD) for a victim transaction trading $25$ ETH to SAI on Uniswap. In contrast, the adversary gains a revenue of $0.14$ ETH ($20.71$ USD), if the victim trades $50$ ETH instead of $25$.

\paragraph{Optimal Adversarial Revenue}
\begin{figure}[htb!]
\begin{center}
\includegraphics[width = \columnwidth]{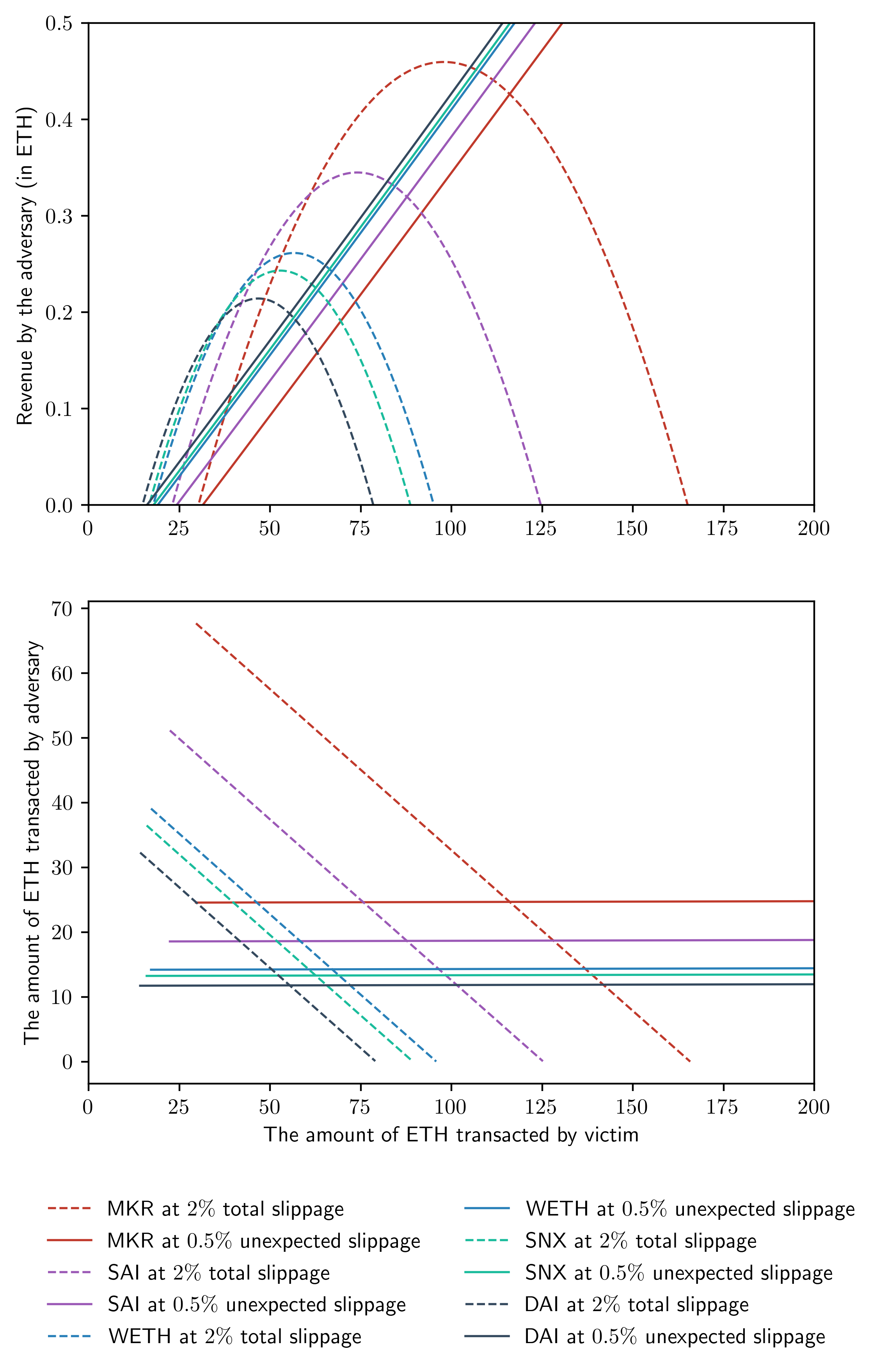}
\end{center}
\caption{Sandwich attack optimal revenue for an adversarial taker when $V$ trades on five Uniswap markets ($0.3\%$ fee, $A$ breaks even at $0.01$ ETH).}
\label{fig:optimal-attack1}
\end{figure}

\begin{figure}[htb!]
\begin{center}
\includegraphics[width = \columnwidth]{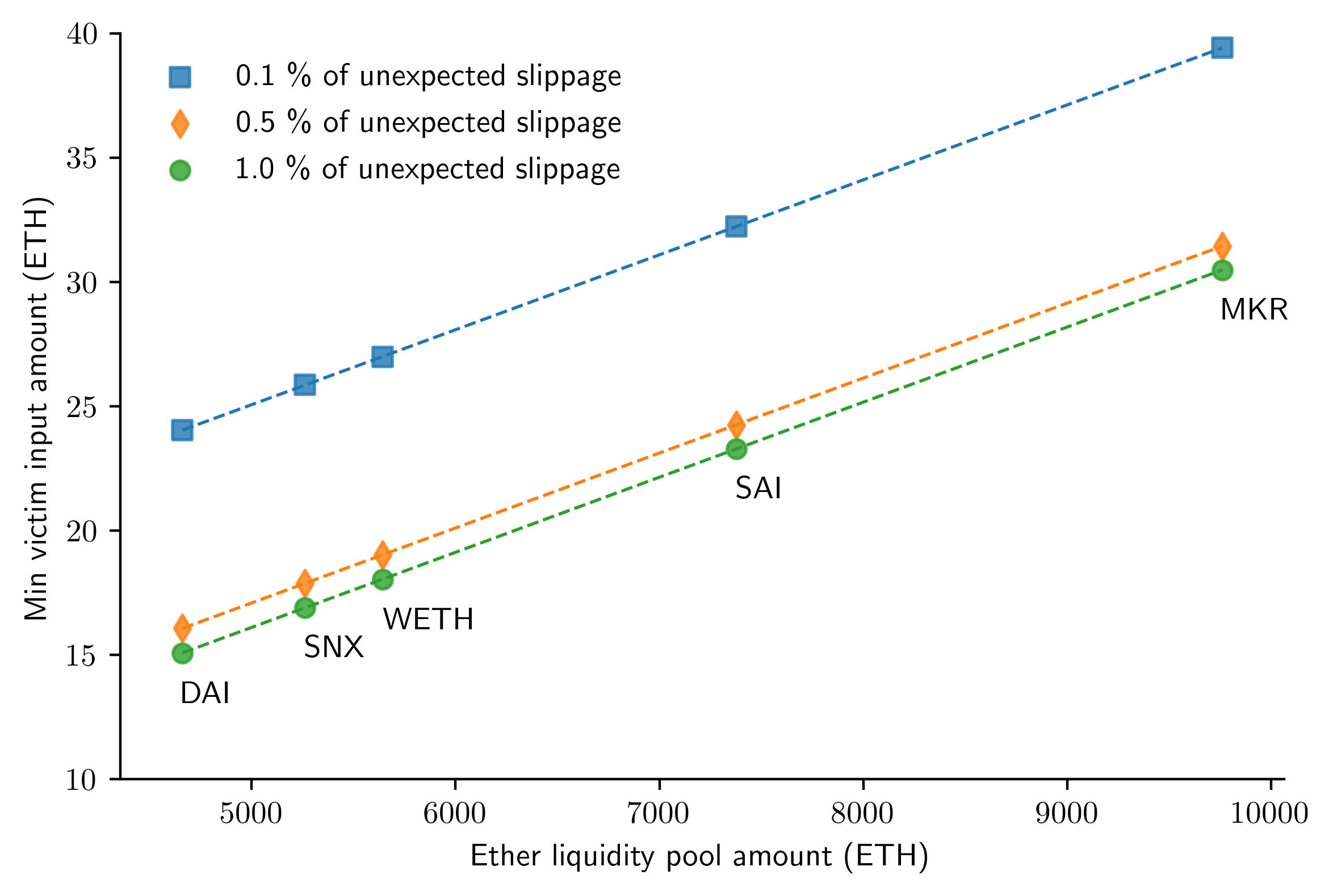}
\end{center}
\caption{Minimum profitable victim input on five Uniswap markets ($0.3\%$ fee, $A$ breaks even at $0.01$ ETH, $0.5\%$ unexpected slippage). A liquidity taker does not yield a profit if $T_V$ trades less than $24.26$ ETH for SAI.}
\label{fig:minimum_victim_input_attack1}
\end{figure}

Out of the over $1,300$ Uniswap exchange markets (i.e.\ coin pairs to trade) an adversary may need to focus and hold liquidity in selected markets. 
In Figure~\ref{fig:optimal-attack1} we quantify the maximum revenue an adversary can expect in a given market, conditional on a suitable victim transaction. 
Note that MKR has the highest liquidity ($9,759.83$ ETH and $2,830.27$ MKR), followed by SAI ($7,377.53$ ETH and $1,099,040.91$ SAI), WETH ($5,642.08$ ETH and $5,632.25$ WETH), SNX ($5,262.53$ ETH and $572,512.14$ SNX) and DAI ($4,660.75$ ETH and $693,706.47$ DAI).

\paragraph{Minimum Profitable Victim Input}
Not every victim transaction yields a profitable attack. For each of the five exchanges in Figure~\ref{fig:optimal-attack1}, we quantify a minimum profitable victim input $\text{min.}_{input}$ (under $0.01$ ETH transaction fee and $0.3\%$ commission), under which an adversary will be unable to make a profit (e.g.\ $24.26$ ETH for SAI per Figure~\ref{fig:optimal-attack1}). This minimum profitable victim input amount increases with the liquidity pool size (cf.\ Figure~\ref{fig:minimum_victim_input_attack1}). The adversary's optimal input increases only slightly (cf.\ the near horizontal line on Figure~\ref{fig:optimal-attack1}) with the victim transaction size, because the ETH value transacted by the victim is relatively small compared to the total amount of ETH in the Uniswap exchange. Given a fixed total slippage, we observe that markets with higher liquidity (e.g.\ SAI, MKR) yield higher potential revenues than lower-liquidity markets (e.g.\ SNX and DAI) (given the appropriate victim transaction).

\subsection{Adversarial Liquidity Provider} 

\begin{figure*}[h!]
\centering
\subfigure[ETH/SAI market]{
\label{fig:revenue_attack_2_SAI}
\includegraphics[height=2.8in]{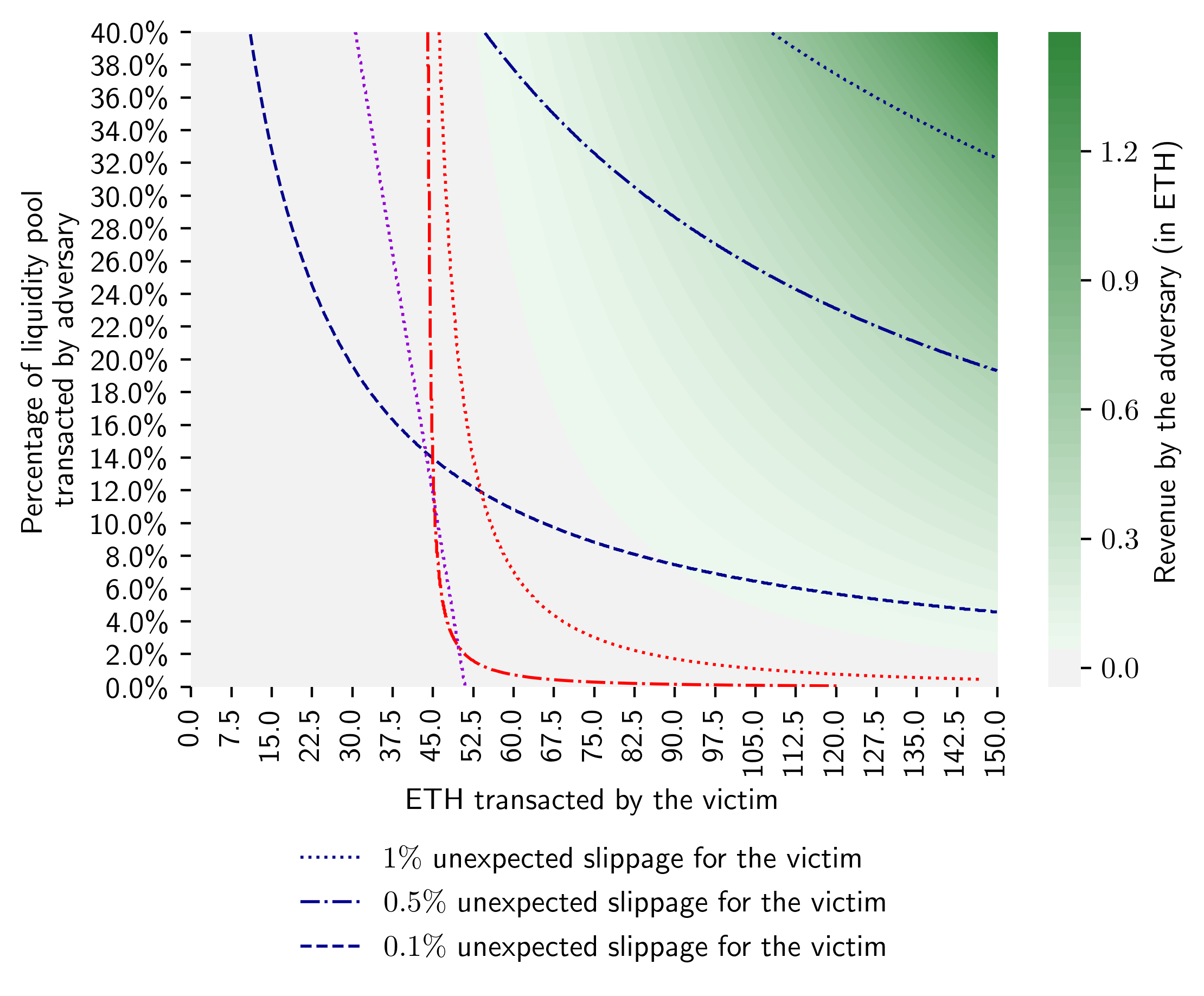}}%
\qquad
\subfigure[ETH/DAI market]{
\label{fig:revenue_attack_2_DAI}
\includegraphics[height=2.8in]{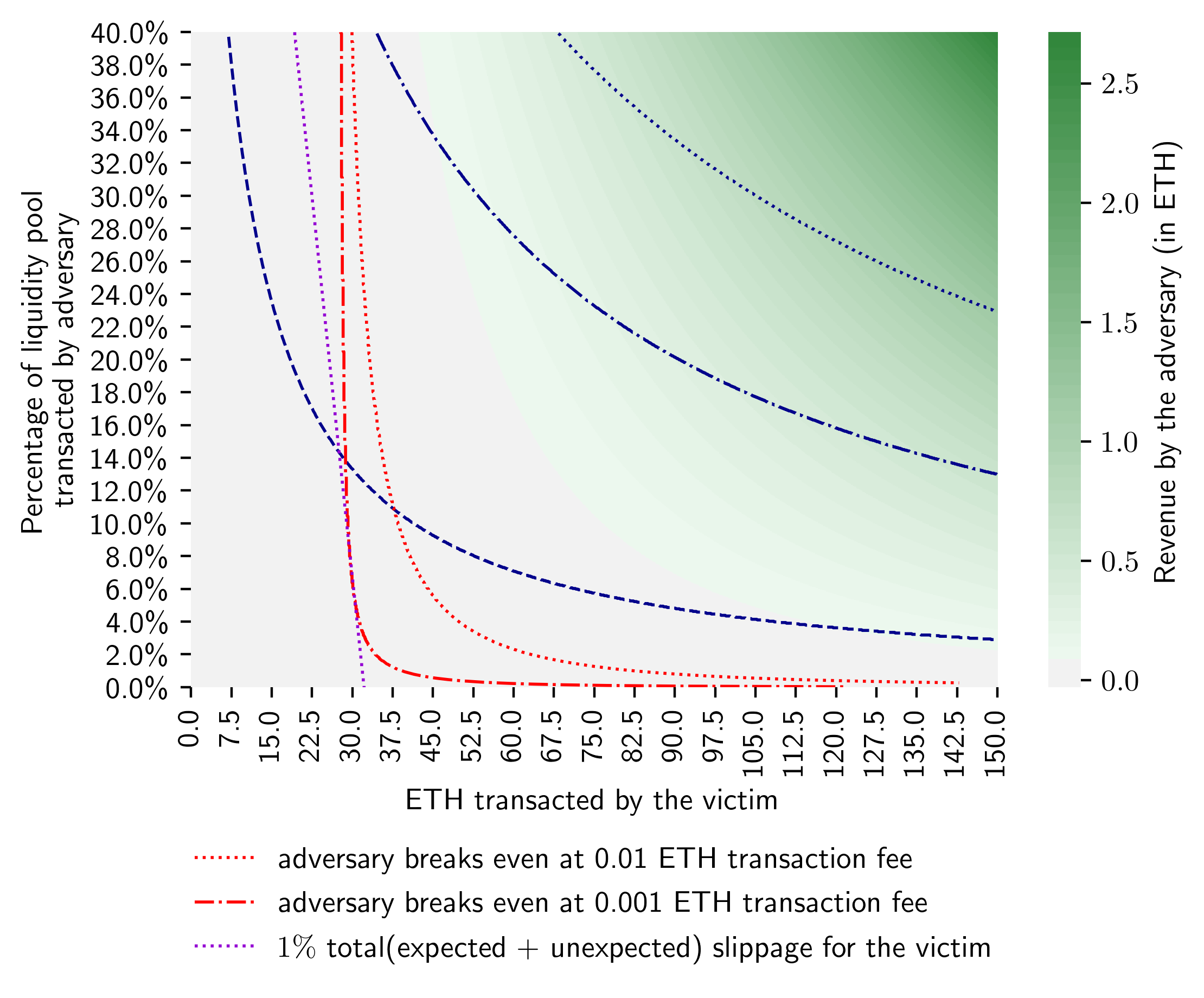}}
\caption{Sandwich attack by a liquidity provider on a taker (Uniswap, block $9$M, $0.3\%$ fees). If $T_V$ trades SAI for $60$ ETH with an unexpected slippage of $0.5\%$, $A$ can achieve a max.\ revenue by front-running $T_V$ with removing $37.76\%$ of liquidity (eq.\ $2,785.97$ ETH and $415,030.47$ SAI), and regain $37.76\%$ of liquidity (deposit $2,749.57$ ETH and $420,542.21$ SAI) by back-running $T_V$. Upon rebalancing to ETH, $A$ gains a profit of $0.07$ ETH ($10.55$ USD, break-even $0.01$ ETH).}
\end{figure*}

\begin{figure}[htb!]
\begin{center}
\includegraphics[width = 0.95\columnwidth]{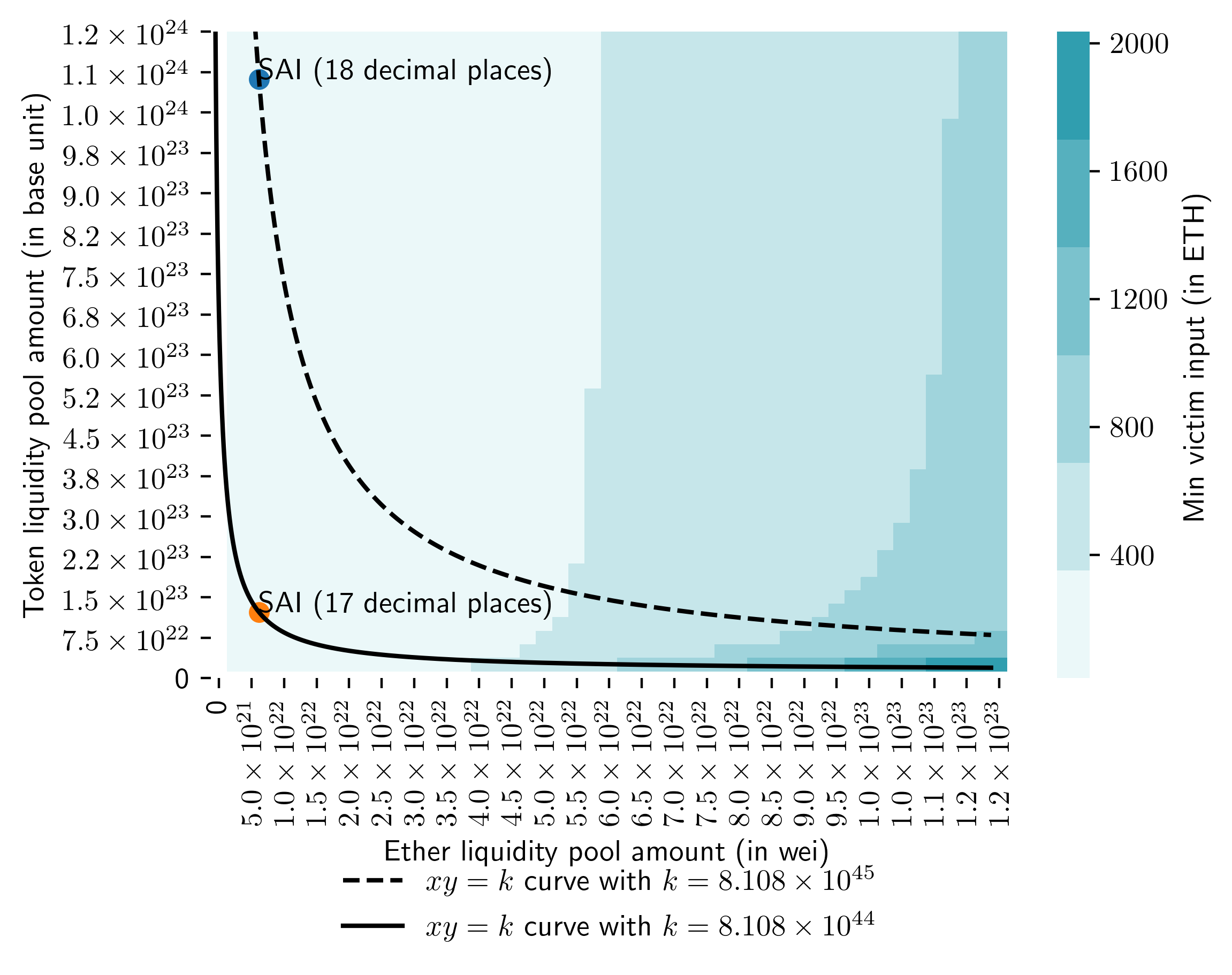}
\end{center}
\caption{Minimum profitable victim input on SAI Uniswap market ($0.3\%$ fee, $0.5\%$ unexpected slippage, adversary break even at $0.01$ ETH tx fees). $A$ cannot gain any profit, if $T_V$ trades SAI for less than $43.93$ ETH. If SAI had $17$ decimal places after the comma instead of $18$, the min.\ victim transaction amount increases to $44.54$ ETH.} 
\label{fig:minimum_victim_input_attack2} 
\end{figure}

Figure~\ref{fig:revenue_attack_2_SAI} and~\ref{fig:revenue_attack_2_DAI} show the revenue of an adversarial liquidity provider (cf.\ Section~\ref{sec:sandwich-attack2}), after $T_V$. We visualize the same adversarial break even lines at $0.01$ ETH ($1.97$ USD) and $0.001$ ETH ($0.2$ USD). Note that the adversary can only withdraw a limited amount of liquidity without triggering the slippage protection on victim's transaction. By removing liquidity from an AMM market, the liquidity provider is forgoing a market commission ($0.3\%$ for Uniswap). To gauge profitability, we consider the following example where $T_V$ purchases $100$ ETH from the SAI Uniswap exchange with $0.5\%$ unexpected slippage. The optimal strategy is to withdraw $26.58\%$ of the total liquidity pool, which leads to a revenue of up to $0.28$ ETH ($41.71$ USD) for the adversary. A passive liquidity provider with $26.58\%$ of the liquidity pool would only earn $0.08$ ETH ($11.91$ USD) given a commission of $0.3\%$.

\paragraph{Who Loses Money?}
Both $V$ and other honest liquidity providers lose money. $V$ purchases $100$ ETH with $15,223.02$ SAI without triggering a $0.5\%$ slippage protection as a result of $A$'s front-running transaction. With no adversary, $V$ only needs $15,147.28$ SAI for the same amount of ETH, which is $75.74$ SAI less. In addition, this $V$ transaction should increase the liquidity pool from ($7,377.53$ ETH / $1,099,040.91$ SAI) to ($7,277.53$ ETH / $1,114,263.92$ SAI). Post attack, the liquidity pool remains with $7,277.25$ ETH and $1,114,263.92$ SAI, i.e\ $A$ gains $7,277.53 - 7,277.25 = 0.28$ ETH from the liquidity pool.

\paragraph{Optimal Adversarial Revenue} 
We quantify the optimal adversarial revenue in Figure~\ref{fig:optimal-attack2}, \emph{after subtracting the foregone opportunity cost} (e.g.\ 0.3\% for liquidity provider on Uniswap), conditional on a suitable victim transaction. We observe that the forgone commission fee is relatively stable, given a fixed total slippage, because the adversary must satisfy the victim's slippage limit. We also quantify in Figure~\ref{fig:minimum_victim_input_attack2} the minimum victim input.

\paragraph{Impact of Coin Decimals}
The number of decimal places for ERC20 tokens is configurable, though most of the coins have 18 decimal places. For example, USDC, which is the $7$th largest Uniswap exchange at block $9$M, has $6$ decimal places behind the comma. In Figure~\ref{fig:minimum_victim_input_attack2}, we plot the $k=xy$ curve for SAI if it had $17$ decimal places instead of $18$. The minimum victim transaction amount for $A$ increases from $43.93$ ETH to $44.54$ ETH, if the victim purchases ETH using SAI. The minimum victim transaction amount also increases from $45.3$ ETH to $56.3$ ETH for ETH to SAI transactions.

\subsection{Overall success of the attacks}
Overall, when analytically evaluated, both an adversarial liquidity taker and provider can profit by undertaking a sandwich attack. The optimal adversarial revenue, however, depends on the slippage protection setting. By fixing the unexpected slippage, the adversary's revenue increases linearly against the amount of ETH transacted for both adversarial takers and providers. Alternatively, fixing the total slippage (unexpected + expected slippage) would yield an upper bound for both the victim transaction size and adversarial optimal profit.

\section{Empirical Evaluation}\label{sec:empirical-evaluation}
Our experimental setup corresponds to the system model in Figure~\ref{fig:system}, with a modified adversarial Parity client. We increase the maximum number of transactions in the pool of unconfirmed transactions (mempool) from the default $1024$ to $2048$. We design a Python script that subscribes to the modified pub/sub functionality of Parity and listens for new pending transactions of the target Uniswap market. Our script computes the profitability of any given victim transaction, and if an adversarial strategy proves profitable, the script generates and propagates the corresponding front- and back-running transaction.

We conduct both experiments (cf.\ Section~\ref{sec:eval-taker} and~\ref{sec:eval-provider}) on the main Ethereum network against the ETH/VERI Uniswap market and only attack our transactions. The ETH/VERI market offers the smallest liquidity ($0.01$ ETH and $0.07$ VERI, total $3.50$ USD) out of the $78$ Uniswap exchanges on the Uniswap UI as of block $9$M. To ensure that our results are sufficiently representative, we consider a time window of 158 days, i.e.\ several months. Our adversarial node runs on AWS in Ireland, ($4$ vCPU, AMD EPYC $7000$, $2.5$ GHz, NVMe SSD, max.\ $10$ Gbps network ). The experiments result in three outcomes: \emph{(i)} success (all adversarial transactions are included in the same block as $T_V$), \emph{(ii)} the front- and back-running transactions are successful, but not all adversarial transactions are included in the same block as $T_V$, and \emph{(iii)} front- or/and back-running failed.

\paragraph{Computing the adversarial transactions}
Three steps allow us to compute the optimal adversarial input amount. First, the maximum amount $A$ can transact without breaking $V$'s slippage protection (denoted by $\hat{o}$) is computed using a binary search. Second, we calculate if the attack is profitable if $A$ inputs $\hat{o}$. As Figure~\ref{fig:SandwichAttackLTvsLT} and \ref{fig:SandwichAttackLPvsLT} suggest, if an attack is not profitable at $\hat{o}$, then it is not profitable for any $o < \hat{o}$. Finally, because Uniswap uses integer divisions, there might exist $o < \hat{o}$, which results in the same or more profit. We perform a ternary search to find the optimal input.

\subsection{Liquidity Taker Attacking Taker}\label{sec:eval-taker}
We issue and attack $T_V$ purchasing VERI with $0.001$ ETH. $T_V$ is triggered through the Uniswap UI (default $0.5\%$ unexpected slippage) and at the time of writing default Metamask gas price ($5$ GWei\footnote{1 GWei = $1 \times 10^{-9}$ ETH}). We repeat this attack $20$ times, and report the results in Table~\ref{tab:experiment}. On average, the adversary discovers $T_V$ within $450$ms, and requires less than $200$ms to compute and send out $T_{A1}$ and $T_{A2}$. During our experiment, $T_V$ remains in the adversary's mempool for an average of $35.84$ seconds. We achieve a success rate of $19$ out of $20$ attempts. One experiment failed, where the victim's transaction $T_V$ remained in the adversarial's mempool for only $1.677$ seconds. In $8$ out of $20$ experiments, the attack is partially successful, because the back-running transaction $T_{A2}$ is mined in a later block than $T_V$. Two possible causes are that either $T_{A2}$ is received after $T_{A1}$ and $T_{V}$ are mined, or the block that mined $T_V$ is full. We observe that the respective $T_V$ are mostly positioned at the end of the block, which may indicate network congestion.

\subsection{Liquidity Provider Attacking Taker}\label{sec:eval-provider}
We initialize the adversary by adding liquidity to the ETH/VERI Uniswap contract. We again issue $T_V$ purchasing VERI with $0.002$ ETH via the Uniswap UI, Metamask ($2$ GWei), and attack with our adversarial node. We also repeat this attack $20$ times. Table~\ref{tab:experiment} shows a summary of our experiment results. Compared to Section~\ref{sec:eval-taker}, $T_V$ remains, on average less than $10$ seconds in the mempool, which may indicate that the blockchain network is less congested at the time of the experiment. We also observe that the adversarial transactions are relatively closer to $T_V$ within the block. We achieve a success rate of $20$ out of $20$ attempts.

\begin{table}[htb!]
\centering
\resizebox{\columnwidth}{!}{
\small
\begin{tabular}{p{4.2cm}p{1cm}p{1cm}p{1cm}p{1cm}}
\toprule
{} & \multicolumn{2}{p{3cm}}{Adversarial Taker} & \multicolumn{2}{p{3.5cm}}{Adversarial Provider} \\
\midrule
{} & Mean & STD & Mean & STD \\
\midrule
$T_{V}$ Broadcast Duration   & $0.45$s  & $0.27$  & $0.36$s  & $0.29$  \\
A Find Strategy Duration     & $0.03$s  & $0.00$  & $0.03$s  & $0.00$  \\
A Execute Strategy Duration  & $0.16$s  & $0.60$  & $0.04$s  & $0.00$  \\
$T_{V}$ Duration In Mempool  & $35.84$s & $33.31$ & $23.09$s & $10.52$ \\
$T_{A1}$ Duration In Mempool & $35.88$s & $33.19$ & $23.03$s & $10.52$ \\
$T_{A2}$ Duration In Mempool & $48.87$s & $51.25$ & $23.03$s & $10.52$ \\
$T_{A3}$ Duration In Mempool & N/A      & N/A     & $23.03$s & $10.52$ \\
A1 Block Relative Position   & $0.05$   & $0.22$  & $0.00$   & $0.00$  \\
A1 Index Relative Position   & $-10.42$ & $10.26$ & $-2.95$  & $2.34$  \\
A2 Block Relative Position   & $0.70$   & $0.92$  & $0.00$   & $0.00$  \\
A2 Index Relative Position   & $5.45$   & $6.47$  & $4.50$   & $4.90$  \\
A3 Block Relative Position   & N/A      & N/A     & $0.00$   & $0.00$  \\
A3 Index Relative Position   & N/A      & N/A     & $5.50$   & $4.90$  \\
\midrule
Success & \multicolumn{2}{l}{$11$/$20$} & \multicolumn{2}{l}{$20$/$20$}\\
Partial Success  & \multicolumn{2}{l}{$8$/$20$}  & \multicolumn{2}{l}{$0$/$20$} \\
Failure          & \multicolumn{2}{l}{$1$/$20$}  & \multicolumn{2}{l}{$0$/$20$} \\
\bottomrule
\end{tabular}
}
\caption{Results for the liquidity taker/provider attacks taker. The victim's transaction $T_V$ was triggered manually using Metamask through Uniswap UI. Adversarial node and victim have a clock difference of $8.781ms \pm 6.189ms$.}
\label{tab:experiment}
\end{table}

\subsubsection{Foregone Adversarial Revenues}
To understand the financial potential of our attacks, we estimated the theoretical revenue for the $79$ exchanges of the Uniswap UI between block $8$M and $9$M (i.e.\ recent blocks at the time of writing), assuming a break-even at $0.01$ ETH transaction fees. Our results (cf.\ Table~\ref{tab:revenue}) suggest that within the reported 158 days, an adversary could have achieved a revenue of $440,749.02$ USD when attacking as a taker, and $98,666.15$ USD when attacking as a liquidity provider. $7.4\%$ of transactions are profitable when attacking as a taker, while $4.2\%$ when attacking as a provider.

\begin{table}[htb!]
\centering
\resizebox{\columnwidth}{!}{
\small
\setlength\tabcolsep{5pt}
\begin{tabular}{@{}p{3cm}p{2cm}p{2cm}p{2cm}@{}}
\toprule
                                      & Profitable~TXs / Total~TXs & Revenue (ETH) & Revenue (USD) \\ 
\midrule
\multicolumn{4}{l}{Liquidity taker attacks taker} \\  
\midrule
ETH $\rightarrow$ Token   & $878$/$25,204$    & $98.15$    & $14,621.41$ \\
Token $\rightarrow$ ETH   & $5,657$/$602,85$  & $2,643.84$ & $393,852.46$ \\
Token $\rightarrow$ Token & $1,258$/$196,72$  & $216.66$   & $32,275.16$ \\
\midrule
Total                     & $7,793$/$105,161$ & $2,958.64$ & $440,749.02$ \\
\midrule
\multicolumn{4}{l}{Liquidity provider attacks taker} \\
\midrule
ETH $\rightarrow$ Token   & $444$/$25,204$   & $52.55$  & $7,829.05$ \\
Token $\rightarrow$ ETH   & $3,254$/$60,285$  & $520.61$ & $77,555.62$ \\
Token $\rightarrow$ Token & $721$/$19,672$   & $89.16$  & $13,281.49$ \\
\midrule
Total                     & $4,419$/$105,161$ & $662.32$ & $98,666.15$ \\
\bottomrule
\end{tabular}
}
\caption{Estimated adversarial revenue for the $79$ exchanges on the Uniswap UI, assuming an adversarial break even cost of $0.01$ ETH. Data of 158 days considered (block $8$M to $9$M).}
\label{tab:revenue}
\end{table}

\subsection{Slippage}
\begin{figure}[htb!]
\begin{center}
\includegraphics[width = 0.95\columnwidth]{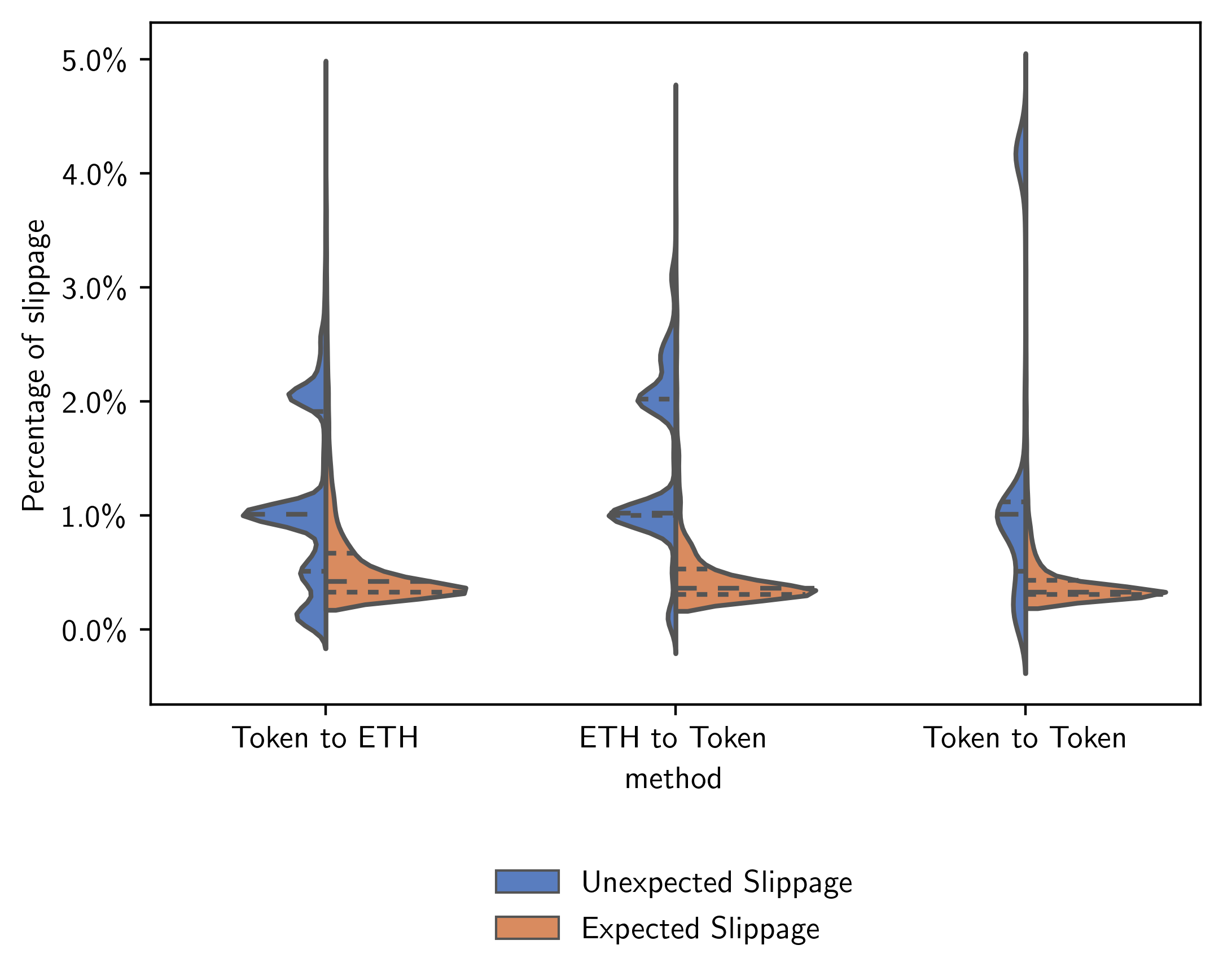}
\end{center}
\caption{Estimated expected and maximum permitted unexpected slippage on Uniswap transactions (block $8$M to $9$M). Most takers trade with c.\ $1\%$ of maximum unexpected slippage (the Uniswap default at the time).}
\label{fig:uniswap-slippage}
\end{figure}

To help an adversary understand how takers configure their slippage, we plot the estimated distribution of expected slippage and maximum allowed unexpected slippage in Figure~\ref{fig:uniswap-slippage}. Note that we are using block $N$'s state to calculate the slippages of a transaction mined at block $N+1$. Therefore, these slippages are only estimates, as we do not know the exact block state used by the taker to create transactions. Past Uniswap transactions have an average expected slippage of $0.58\%$, and an average unexpected slippage of $1.16\%$. 

\subsection{Overall success of the attacks}
Our empirical results suggest that both an adversarial liquidity taker and provider can again profit by undertaking a sandwich attack, where the victim trades with the Uniswap default slippage protection strategy at the time of writing this paper ($0.5\%$ total slippage). We crawled the previous transactions on Uniswap, where it shows that the most common unexpected slippage configuration is $1\%$, which is higher than the $0.5\%$ default total slippage and therefore leads to higher front-running profit. Our experiments result in a high success rate (only 1 out of 40 failed), mainly because the Ethereum network was not congested, and the VERI market relatively inactive.

\section{How Miners Order Transactions}\label{sec:block-position}
One crucial aspect of the potential profitability of the sandwich attacks centers on how miners order transactions within blocks. Blockchains typically prescribe specific rules for consensus, but there are only loose requirements for miners on how to order transactions within a block. To gain insight into this, we crawled the Ethereum blockchain from block $6,627,917$ (where Uniswap was launched) to block $9$M, constituting a total of $2,372,084$ blocks, or equivalently $388$ days of data. For each block, we classified the order of its transactions into one of four classes:
\begin{description}
\item[Empty:] A block without transactions.
\item[Gas Price:] All transactions are sorted in descending order according to the gas price of each transaction.
\item[Parity Default:] Transactions are split into groups according to Parity's prioritization (e.g.\ local transactions first, penalized transactions last). Then, within each group, the transactions are sorted in descending order according to each transaction's gas price.
\item[Unknown:] Transactions are not ordered by the gas price and do not follow parity's default strategy.
\end{description}

\begin{table*}[htb!]
    \centering
    \resizebox{0.9\textwidth}{!}{
    \begin{tabular}{@{}lrrrrr@{}}
    \toprule
     & \multicolumn{4}{c}{Transaction Ordering Strategy} \\ \cmidrule(l){2-5}
    Miner Address & Empty & Gas Price & Parity Default & Unknown & Total Blocks \\ \midrule
    \texttt{\href{https://etherscan.io/address/0xea674fdde714fd979de3edf0f56aa9716b898ec8}{0xea674fdde714fd979de3edf0f56aa9716b898ec8}} & 9,434 & 400,546 & 171,428 & 23,254 & 604,662 \\
    \texttt{\href{https://etherscan.io/address/0x5a0b54d5dc17e0aadc383d2db43b0a0d3e029c4c}{0x5a0b54d5dc17e0aadc383d2db43b0a0d3e029c4c}} & 16,329 & 423,103 & 93,903 & 35,348 & 568,683 \\
    \texttt{\href{https://etherscan.io/address/0x829bd824b016326a401d083b33d092293333a830}{0x829bd824b016326a401d083b33d092293333a830}} & 7,727 & 275,013 & 1,900 & 20 & 284,660 \\
    \texttt{\href{https://etherscan.io/address/0x52bc44d5378309ee2abf1539bf71de1b7d7be3b5}{0x52bc44d5378309ee2abf1539bf71de1b7d7be3b5}} & 259 & 210,061 & 59,141 & 369 & 269,830 \\
    \texttt{\href{https://etherscan.io/address/0xb2930b35844a230f00e51431acae96fe543a0347}{0xb2930b35844a230f00e51431acae96fe543a0347}} & 0 & 110,079 & 19,271 & 125 & 129,475 \\
    \texttt{\href{https://etherscan.io/address/0x04668ec2f57cc15c381b461b9fedab5d451c8f7f}{0x04668ec2f57cc15c381b461b9fedab5d451c8f7f}} & 7,130 & 42,405 & 257 & 0 & 49,792 \\
    \texttt{\href{https://etherscan.io/address/0x2a65aca4d5fc5b5c859090a6c34d164135398226}{0x2a65aca4d5fc5b5c859090a6c34d164135398226}} & 1,021 & 25,569 & 10,892 & 251 & 37,733 \\
    \texttt{\href{https://etherscan.io/address/0x2a5994b501e6a560e727b6c2de5d856396aadd38}{0x2a5994b501e6a560e727b6c2de5d856396aadd38}} & 1,170 & 31,543 & 5 & 0 & 32,718 \\
    \texttt{\href{https://etherscan.io/address/0x005e288d713a5fb3d7c9cf1b43810a98688c7223}{0x005e288d713a5fb3d7c9cf1b43810a98688c7223}} & 1,097 & 27,926 & 30 & 0 & 29,053 \\
    \texttt{\href{https://etherscan.io/address/0x35f61dfb08ada13eba64bf156b80df3d5b3a738d}{0x35f61dfb08ada13eba64bf156b80df3d5b3a738d}} & 435 & 28,214 & 68 & 0 & 28,717 \\ \bottomrule
    \end{tabular}
    }
    \caption{Classification of the top 10 miners in Ethereum, in terms of the number of blocks mined between blocks $6,627,917$ until $9$M ($388$ days). We see that miners seem to switch among strategies. Moreover, 4 out of the ten miners always seem to follow a known strategy. They either order their transactions by gas price or by using Parity's default strategy. We also note that the address \texttt{0xb293..0347} is the sole miner who did not mine any empty blocks.}
    \label{tab:top-miner-tx-order}
\end{table*}

The treatment of transactions depends on the Ethereum client. At of the time of writing, $78.3\%$ of the Ethereum clients operate Geth, respectively $20.2\%$ Parity\footnote{\url{https://www.ethernodes.org/}}. Geth first sorts and separates the list of transactions into lists of individual sender accounts and sorts them by nonce. Afterward, they are merged back together and sorted by gas price, always comparing only the first transaction from each account. Parity, by default, prioritizes local and retracted transactions first, and polishes transactions with heavy computation, and then sorts by gas price. A transaction is considered local if it is received via the RPC interface, or the sender of the transaction is part of the list of locally managed accounts. Miners may choose to modify the transaction inclusion policy arbitrarily. To individually categorize each block, we first analyzed the gas price of each transaction and started by extracting the gas price only for the first transaction of each sender, while ignoring the other transactions from the same sender. We only consider the first transaction because a higher gas price transaction can be placed behind another transaction with a smaller nonce. If the extracted gas price list is sorted, we classify the block as following the ``gas price'' strategy. Alternatively, we verify if the gas price list consists of multiple sublists of gas prices, where the gas prices within each sublist are sorted in descending order. Each sublist represents a priority group, where transactions within the same priority group are sorted by gas price. If there are no more than four sublists of gas prices (local, retracted, normal, penalized), we classify the block as a ``Parity default'' block, otherwise, as ``unknown''.

Because both Parity and Geth sort transactions by default by gas price, it is difficult to identify which client a miner uses. Moreover, our heuristics may misclassify blocks as being ordered using gas price instead of Parity's default strategy. A miner could have no local transactions for the block, and all transactions are thus ordered by gas price. The number of blocks classified as Parity default should, therefore, be regarded as a lower bound. We leave it for future work to develop a more precise client fingerprinting strategy.

\subsection{Miner Transaction Ordering Results}
Our results (cf.\ Table~\ref{tab:overall-tx-order} and Table~\ref{tab:top-miner-tx-order} suggest that roughly 79\% of the miners order transactions based on the gas price, thus likely following Geth's strategy, and 16\% order their transaction after Parity's default strategy. These results are consistent with the client usage distribution. We find that 2\% of the blocks are empty. 3\% of the miners follow an unknown ordering method. We can conclude that to position a transaction \emph{before} or \emph{after} a target transaction, it is, with a probability of $79$\%, sufficient to pay a higher ($+1$ Wei) or lower gas price ($-1$ Wei) --- assuming the absence of other front-running adversaries. If two transactions pay the same gas price, according to their source code, Parity and Geth include the transactions after the FIFO principle. 


Overall, we find that most miners (c.\ 79\%) order transactions based on the gas price. The dominance of this transaction order aids an adversary because it makes it more likely for an adversary that they can control the placement of their transactions in a block by tweaking the gas price they offer for each transaction.

\begin{table}[htb!]
\centering
\begin{tabular}{l r r}
\toprule
Strategy & Number of Blocks & Ratio \\ 
\midrule
Empty Block 	 	        & $55,545$ 	& $0.0234$ \\
Order per Gas Price	 	    & $1,862,800$ & $0.7853$ \\
Order per Parity Default    & $384,150$   & $0.1620$ \\
Unknown Ordering            & $69,589$    & $0.0293$ \\
\midrule
Total                       & $2,372,084$ & $1.0000$ \\ \bottomrule
\end{tabular}
\caption{Miner transaction ordering methodology between blocks $6,627,917$ and $9$M ($388$ days).}
\label{tab:overall-tx-order}
\end{table}

\subsection{Gas Price Distribution}
We investigate the gas prices for each transaction over $189,951,899$ transactions included in blocks $6,627,917$ until block $9$M ($388$ days of data). 
A gas price of $0$ ETH might indicate that the transaction belonged to the miner.  We find that a transaction has an average gas price of $17.2 \pm 10520.1$ GWei. The median gas price is $10$ Gwei and the most frequently used gas price is $20$ Gwei with $23,759,990$ transactions ($12.5$\%). 

\section{Multiple Adversaries}\label{sec:multiple-attacker}
Our prior analysis only considers the case of a single adversary. In this section, we analyze the possible implications of multiple attackers on the feasibility and profitability of sandwich attacks through simulations.

\begin{table*}[htb!]
\centering
\begin{tabular}{cccccccc}
\toprule
\multicolumn{5}{c}{Transaction Execution Order (left to right)}         & Winner & Reward for Attacker $A$                                                & Reward for Attacker $O$                                  \\
\midrule
$T_{A1}$ \cmark & $T_{O1}$ \xmark & $T_{V}$ \cmark & $T_{O2}$ \xmark & $T_{A2}$ \cmark & $A$ & $\text{Revenue} - \text{Fee}(T_{A1}) - \text{Fee}(T_{A2})$ & $ - \text{Fee}(T_{O1}) - \text{Fee}(T_{O2})$ \\
\midrule
$T_{A1}$ \cmark & $T_{O1}$ \xmark & $T_{V}$ \cmark & $T_{A2}$ \cmark & $T_{O2}$ \xmark & $A$ & $\text{Revenue} - \text{Fee}(T_{A1}) - \text{Fee}(T_{A2})$ & $ - \text{Fee}(T_{O1}) - \text{Fee}(T_{O2})$ \\
\midrule
$T_{O1}$ \cmark & $T_{A1}$ \xmark & $T_{V}$ \cmark & $T_{A2}$ \xmark & $T_{O2}$ \cmark & $O$ & $ - \text{Fee}(T_{A1}) - \text{Fee}(T_{A2})$ & $\text{Revenue} - \text{Fee}(T_{O1}) - \text{Fee}(T_{O2})$ \\
\midrule
$T_{O1}$ \cmark & $T_{A1}$ \xmark & $T_{V}$ \cmark & $T_{O2}$ \cmark & $T_{A2}$ \xmark & $O$ & $ - \text{Fee}(T_{A1}) - \text{Fee}(T_{A2})$ & $\text{Revenue} - \text{Fee}(T_{O1}) - \text{Fee}(T_{O2})$ \\
\bottomrule
\end{tabular}
\caption{Adversarial payoff for a two player game. Under the assumption that both players $A$ and $O$ are rational, $A$ ``wins'' the game if the front-running transaction $T_{A1}$ is placed in front of $T_{O1}$, regardless of the position of the back-running transaction. A transaction that succeeds is denoted by \cmark, while a transaction that fails, is denoted by \xmark.}
\label{tab:transaction-strategy}
\end{table*}

\subsection{Extended Threat Model}
We extend our threat model from Section~\ref{sec:sandwich-attack}, to account for additional attackers. For simplicity, we assume that all adversaries have access to the same financial resources, internet connection, latency, and computational power. We identify the following key parameters that impact the outcome of the simulated game:
\begin{description}
    \item[Number of Attackers:] Intuitively, the more adversaries are attacking a victim transaction $T_V$, the harder this endeavor becomes for each attacker. In the following, we consider $2$, $5$, and $10$ adversaries, which simultaneously attack a $T_V$.
    \item[Attacker Strategy:] Previous work~\cite{daian2019flash} suggests two transaction fee bidding strategies, namely an adaptive (reactive counter-bidding) and a non-adaptive (blind raising) adversarial strategy. We assume that all adversaries follow the reactive counter-bidding strategy, i.e.\ an adversary emits a higher transaction fee bid once the adversary observes a competing transaction. This strategy is not necessarily optimal, but it may estimate a lower bound for the sandwich attack's front-running transaction cost.
\end{description}
We assume all adversaries are rational and attack with the parameters determined by the strategy from Section~\ref{sec:empirical-evaluation}, i.e.\ each adversary attempts to maximize its profit by fully exploiting the victim transaction's allowed slippage.

For a two-player game, we show in Table~\ref{tab:transaction-strategy}, the possible transaction permutations after the adversarial transactions are mined. We show that the adversary who manages to execute the first front-running transaction successfully ``wins'' the sandwich attack. This is because the victim transaction fails if both $A$ and $O$ execute the sandwich attack. If the other adversary is irrational and insists to execute the attack (by e.g.\ disregarding slippage protections), both adversaries lose, because both adversarial front-running transactions fail (e.g.\ $T_{A1}$ \cmark, $T_{O1}$ \cmark, $T_{V}$ \xmark, $T_{O2}$ \cmark, $T_{A2}$ \cmark).

\subsection{Extended System Model}

\paragraph{Network Layer}
The speed at which an adversarial transaction propagates within the blockchain P2P network influences the number of reactive counter-bids it receives and the time the transaction is mined. Related works have extensively studied the asynchronous nature of blockchain P2P propagation~\cite{ersoy2018transaction, decker2013information, gervais2016security, gencer2018decentralization}. The propagation is affected by several factors, such as the network topology, number of nodes, internet latency, bandwidth, and network congestion, etc. In our work, we assume that the adversary directly establishes a point-to-point connection with the miner and the victim. Our study thereby abstracts away the number of nodes in the network, the network topology, intermediate devices (replay nodes, routes, and switches) and TCP congestion control. Equation~\ref{eq:propagation} shows how we approximate transaction propagation duration.

\begin{equation}\label{eq:propagation}
    \text{Propagation Duration} = \frac{\text{Transaction Size}}{\text{Bandwidth}} + \text{Latency}
\end{equation}

To determine the distribution of transaction sizes, we crawl raw transactions sent to the Uniswap DAI market over $100,000$ consecutive blocks, starting from block 9M. Our measurements suggest a mean transaction size of $426.27 \pm 68.94$ Bytes, which we use as parameters for an assumed normal distribution of the adversarial and victims' transaction sizes. For the latency and bandwidth distribution, we take the mean percentile statistics~\cite{kim2018measuring, gencer2018decentralization} and apply linear interpolation to estimate the underlying cumulative probability distribution (cf. Table~\ref{tab:network_statistics}).

\begin{table}[tb!]
\centering
\begin{tabular}{cccccc}
\toprule
\multirow{2}{*}{Pct \%} & \multicolumn{3}{l}{Latency} & \multicolumn{2}{l}{Provisioned Bandwidth} \\
                        & \cite{kim2018measuring}  & \cite{gencer2018decentralization}       & Model   &   \cite{gencer2018decentralization}                 & Model             \\
\midrule
10                      & 99    & 92    & 95.5        & 3.4               & 3.4                   \\
20                      & 116   &   -   & 116         & -                 & 6.8                   \\
33                      & 151   & 125   & 138         & 11.2              & 11.2                  \\
50                      & 208   & 152   & 180         & 29.4              & 29.4                  \\
67                      & 231   & 200   & 216         & 68.3              & 68.3                  \\
80                      & 247   & -     & 247         & -                 & 111.3                 \\
90                      & 285   & 276   & 281         & 144.4             & 144.4                 \\ \midrule
mean                    & 209   & 171   & 181         & 55.0              & 52.8                  \\
std.                    & 157   & 76    & 62          & 58.8              & 50.4                  \\
\bottomrule
\end{tabular}
\caption{Latency and bandwidth statistics from our model based on previous studies~\cite{kim2018measuring, gencer2018decentralization}.}
\label{tab:network_statistics}
\end{table}

\paragraph{Transaction Fees}
The transaction gas price, together with the degree of blockchain transaction congestion (i.e.\ competing transactions that seek to be mined), influences the pace at which a transaction is mined. In our simulations, we sample the gas price of the victim transaction from a normal distribution with a mean of ($8.76 \pm 61.18$ GWei), measured at the Uniswap DAI market from block $9$M to $9.1$M. We assume that the victim pays a sufficient transaction fee for its transaction to be mined in the next block. Empirical data suggests that the Ethereum average block interval time is $13.5 \pm 0.12$ seconds~\cite{ethereum-block-time}. Therefore, we sample the duration of the victim's transaction in the mempool from a uniform distribution between $0$ to $30$ seconds. When the network is congested, transactions on Ethereum may stay in the mempool for longer than $30$ seconds, sometimes even tens of minutes~\cite{ethereum-block-time}. However, we avoid presenting our analysis on longer pending duration, as our simulation results (cf. Figure~\ref{fig:taker_multiple_adversary} and Figure~\ref{fig:provider_multiple_adversary}) show that in the case of multiple players, the adversarial transaction fee at $30$ seconds is likely to render the attack unprofitable.

\paragraph{Miner}
We assume the miner order transactions with descending gas prices to maximize their revenue (cf. Section~\ref{sec:block-position}). Besides, the miners configure a price bump percentage of $10\%$ to replace an existing transaction from the mempool. At the time of writing, Geth (used by 78.3\% of the clients) configures price bump percentage to $10\%$, while Parity sets $12.5\%$. 

\subsection{Simulation Results}
\begin{figure*}[htb!]
\centering
\subfigure[Two liquidity taker adversaries]{\label{fig:taker_2_adversaries}\includegraphics[width=0.32\textwidth]{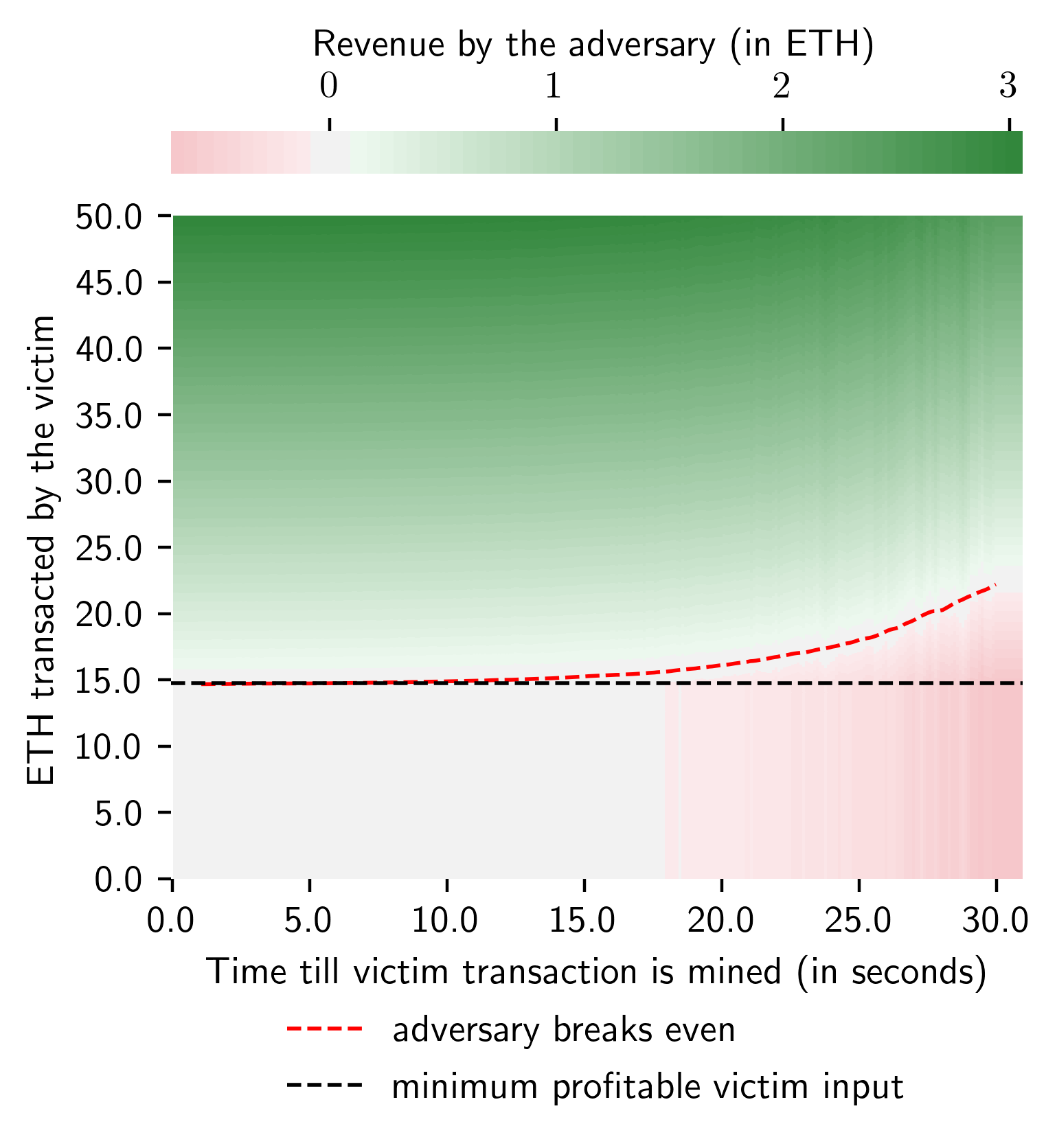}}
\subfigure[Five liquidity taker adversaries]{\label{fig:taker_5_adversaries}\includegraphics[width=0.32\textwidth]{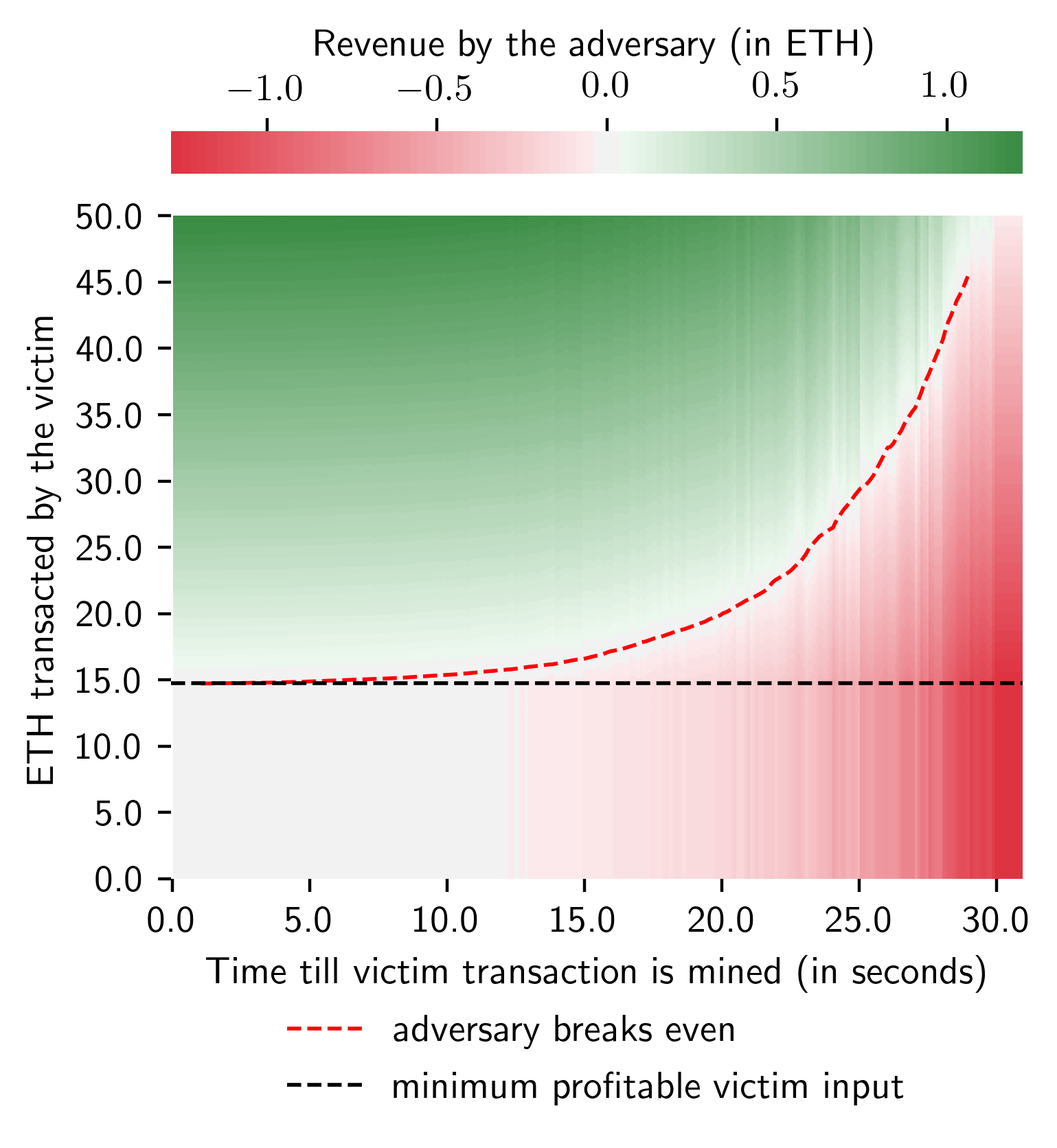}}
\subfigure[Ten liquidity taker adversaries]{\label{fig:taker_10_adversaries}\includegraphics[width=0.32\textwidth]{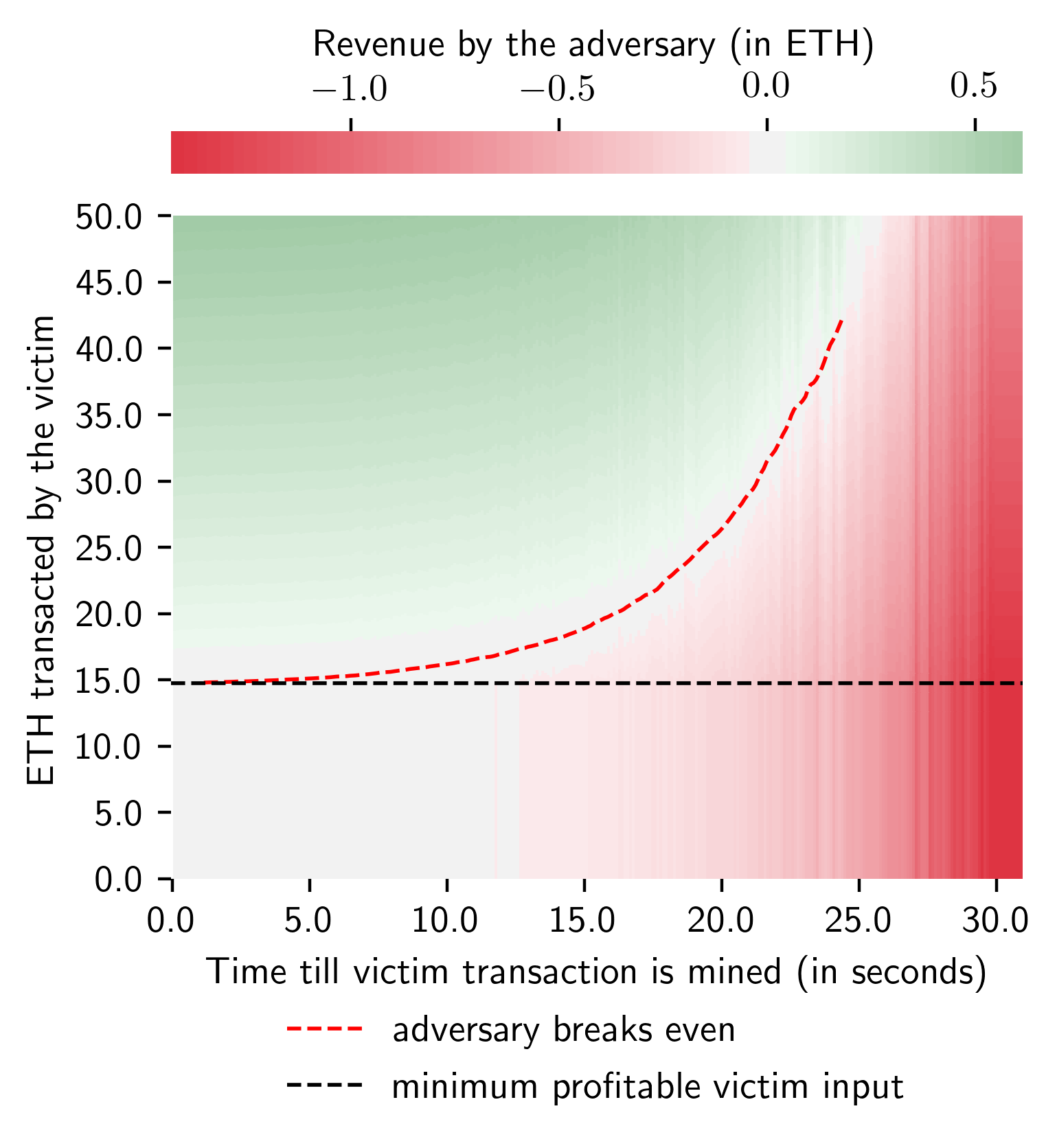}}
\caption{Simulated sandwich attack by $2$, $5$ and $10$ competing adversarial liquidity taker on a taker (Uniswap, block $9$M, $0.3\%$ fees, $0.5\%$ unexpected slippage).}
\label{fig:taker_multiple_adversary}
\end{figure*}

Figure~\ref{fig:taker_multiple_adversary} shows the expected profit of an adversarial liquidity taker (cf.\ Section~\ref{sec:analytical-evaluation}) given $2$, $5$, and $10$ adversaries, on the Uniswap DAI market at block $9$M. The slippage of the victim transaction is fixed at $0.5\%$. The minimum profitable victim input is $14.75$ ETH ($2,197.30$ USD).

\begin{figure}[tb!]
\centering
\includegraphics[width = 0.95\linewidth]{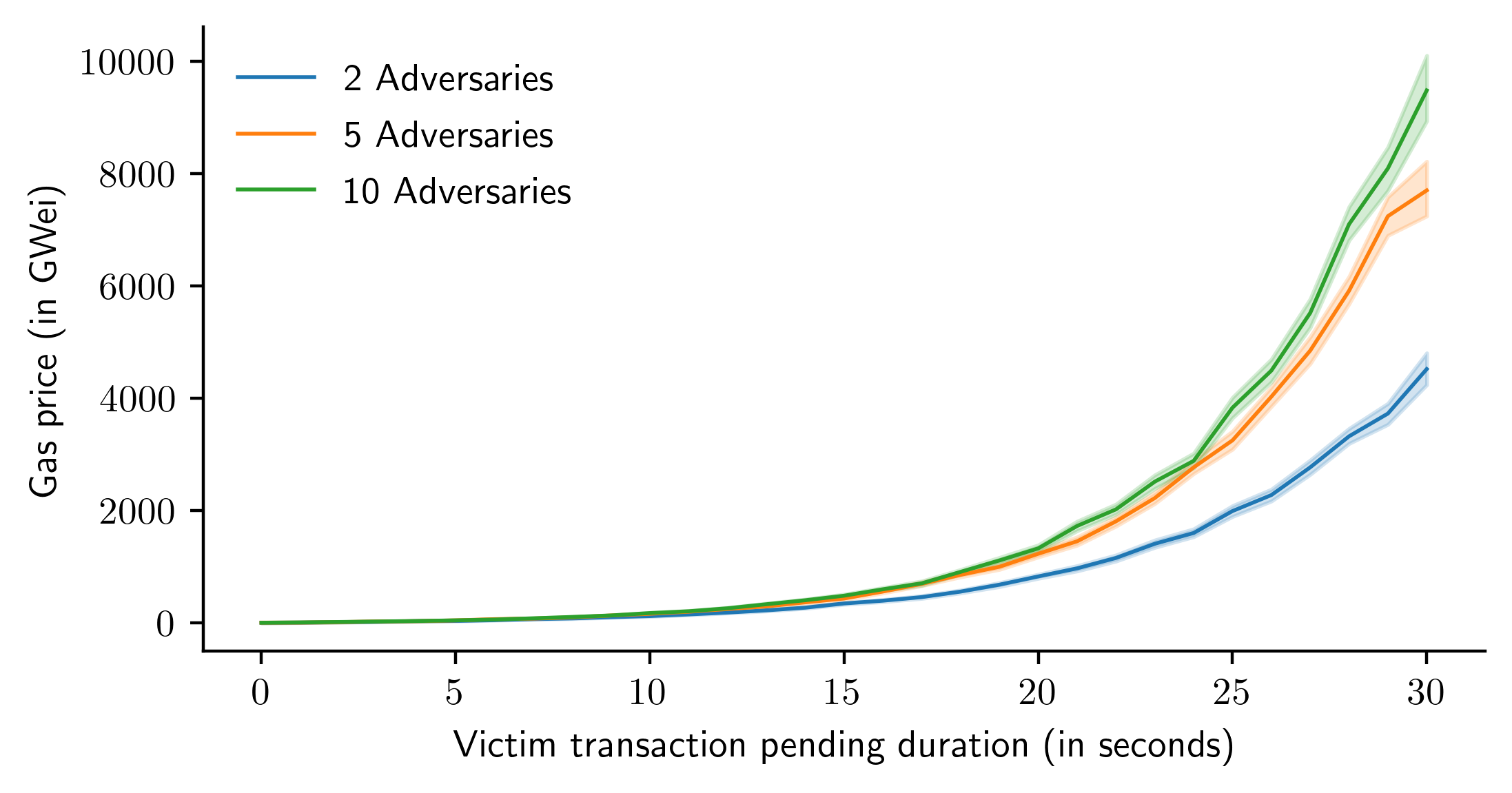}
\caption{The simulated gas price of the ``winning'' transaction when $2$, $5$ or $10$ adversaries are performing a reactive counter-bidding attack. Each experiment is run $100,000$ times. We visualize the $95\%$ confidence interval.}
\label{fig:gas_price_simulation}
\end{figure}

We visualize the line where the expected revenue breaks even with simulated transaction costs. The transaction cost is calculated as the simulated gas price (cf.\ Figure~\ref{fig:gas_price_simulation}) times the total gas consumed by the adversarial transaction. The gas consumption is randomly sampled from a normal distribution with mean at $85,488 \pm 34,782$ (taken from the Uniswap DAI market average gas consumption from block $9$M to $9.1$M). We observe that the break-even line grows exponentially until the victim transaction is mined (which is in line with the assumption of a $10\%$ increase for each transaction price increase). We observe that the more adversaries, the more competitive the attack appears. For instance, our simulation suggests that the sandwich attack is not profitable after the victim transaction remained pending on the P2P network for $27.7$, $20.3$, $16.3$ seconds, given $2$, $5$ and $10$ adversaries respectively, when the victim $V$ transacts $20$ ETH for DAI. We refer the interested reader to Figure~\ref{fig:provider_multiple_adversary} for a visualization of our simulation when multiple providers attack a taker.

Our results suggest that having multiple attackers does, in expectation, divide the total revenue among the adversaries, minus the transaction fee overhead. Specifically, we find that the presence of $2$, $5$ and $10$ attackers respectively reduce the expected profitability of the attack by $51.0$\%, $81.4$\% and $91.5$\% (given the victim transacts $20$ ETH to DAI on Uniswap with a transaction pending on the P2P layer for $10$ seconds before being mined). Note that if the blockchain is congested (i.e.\ $T_V$ remains pending for more than $15$ seconds), we observe that the break-even of the attacker becomes harder to attain.

\section{Related Work}\label{sec:relatedwork}
Besides AMM DEX, other types of decentralised exchanges exist: limit order book based~\cite{kyber2019,oasis2019,idex2019}, auctions~\cite{dutchx2019}, trusted hardware~\cite{bentov2019tesseract}, payment channel~\cite{luo2019payment}. Front-running, and high frequency trading is related to the thoroughly studied problem of rushing adversaries to double-spend not yet mined blockchain transactions~\cite{karame2012double}. Strategically placed and malicious blockchain network nodes may control when and if miners receive transactions, which can affect the time at which a transaction is executed within the blockchain~\cite{marcus2018low, henningsen2019eclipsing, gervais2015tampering, heilman2015eclipse}. The cryptographic literature captures front-running by allowing a ``rushing adversary'' to interact with a protocol~\cite{beaver1992cryptographic}. The (financial) high-frequency trading (HFT) literature~\cite{angel2013fairness, menkveld2016economics} has also explored many trading strategies and their economic impact, such as arbitrage, news reaction strategies etc in traditional markets. Most of the traditional market strategies are also applicable to AMM and other decentralized exchanges~\cite{daian2019flash, angeris2019analysis}. Daian~\emph{et al.}~\cite{daian2019flash} in particular, introduced the concept of gas price auctions (PGA) among trading bots as well as the concept of miner extractable value (MEV). Previous studies~\cite{baron2019risk} also suggest that HFT performance is strongly associated with latency and execution speed. Multiple forms of malpractice have been discovered on financial exchanges. Besides the traditional market manipulation techniques~\cite{jarrow1992market} (such as cornering, front-running, and pump-and-dump schemes), previous works~\cite{lin2016new} have also studied new techniques such as spoofing, pinging and mass misinformation, which leverage modern technologies such as social media and artificial intelligence. Such techniques may even be used to trick HFT algorithms~\cite{arnoldi2016computer}. To counterbalance this inherent trust, regulators conduct periodic and costly manual audits of banks, brokers, and exchanges to unveil potential misbehavior. This is a challenging task on DEX, given weak identities, and missing regulations.

\section{Conclusion}\label{sec:conclusion}
In this paper, we have presented two versions of a sandwich attack, made possible by the deterministic nature of an AMM DEX in combination with the time delay inherent on on-chain exchanges. While the transparency of DEXs is desirable, it can, however, put users assets at a security risk and allow both liquidity providers and liquidity takers to exploit unknowing traders through a combination of front and back-running. Fixing such front-running is not trivial because the smaller the allowed slippage set by a trader, the more likely the trade fails. Cryptography-based defenses moreover affect the usability of the AMM DEXes due to multiple rounds of interactions or trusted off-chain components (cf.\ Appendix~\ref{sec:defences}).

We show how under multiple competing adversaries, sandwich attacks may still remain profitable. Our work, sheds light on a dilemma facing DEXs: if the default slippage is set too low, the DEX is not scalable (i.e.\ only supports few trades per block), if the default slippage is too high, adversaries can profit. We hope that this work draws attention to this unsolved issue and engenders future work on open, secure and decentralized finance.

\section*{Acknowledgments}
We very much thank the anonymous reviewers and Nicolas Christin for the thorough reviews and helpful suggestions that significantly strengthened the paper.


\clearpage

\bibliographystyle{plain}
\bibliography{references.bib}





\appendices

\section{Sandwich Attack Details}
In what follows, we assume that a victim $V$ sends a transaction $T_V$ to trade $\delta_x^V$ of $X$ for $\delta_y^V$ of $Y$. The underlying AMM market starts with the initials state $s_0 = (x_0, y_0)$. Eq.~\ref{eq:initial_state} shows the transition in AMM state after applying $T_V$ without unexpected slippage.

\begin{equation}\label{eq:initial_state}
s_0\xrightarrow{\text{TransactXForY}(\delta_x^V)} (x^*, y^*),\quad \delta_y^V = y_0 - y^*
\end{equation}
\begin{equation}
x^* = x_0 + \delta_x^V,\quad y^* = \frac{x_0 y_0}{x_0 + \delta_x^V - c_x(s_0, \delta_x^V)} + c_y(s_0, \delta_y^V)
\end{equation}

\subsection{Liquidity Taker Attacks Taker}\label{app:taker-attacks-taker}

We assume that the adversary $A$ has an initial state $(\delta_x^{A1}, 0)$, and attacks by emitting $T_{A1}$ and $T_{A2}$.

\vspace{3mm}
\begin{description}
    \item[$T_{A1}$:] a front-running transaction, exchanges $\delta_x^{A1}$ for $\delta_y^{A1}$ of $Y$ and is planned to execute before $T_V$ (e.g.\ by paying a higher transaction fee than $T_V$). 
    This results in the state changes of Equation~\ref{eq:lt-a1-1} and~\ref{eq:lt-a1-2}.
    \begin{equation}\label{eq:lt-a1-1}
    s_0\xrightarrow{\text{TransactXForY}(\delta_x^{A1})} s_1=(x_1, y_1),\quad \delta_y^{A1} = y_0 - y_1
    \end{equation}
    \begin{equation}\label{eq:lt-a1-2}
    x_1 = x_0 + \delta_x^{A1},\quad y_1 = \frac{x_0 y_0}{x_1 - c_x(s_0, \delta_x^{A1})} + c_y(s_0, \delta_x^{A1})
    \end{equation}
    
    \item[$T_V$:] a victim transaction modifies the state as per Equation~\ref{eq:lt-victim1} and~\ref{eq:lt-victim2}.
    \begin{equation}\label{eq:lt-victim1}
    s_1\xrightarrow{\text{TransactXForY}(\delta_x^{V})} s_2=(x_2, y_2),\quad \delta_y^{V} = y_1 - y_2
    \end{equation}
    \begin{equation}\label{eq:lt-victim2}
    x_2 = x_1 + \delta_x^{V},\quad y_2 = \frac{x_1 y_1}{x_2 -  c_x(s_1, \delta_x^{V})} + c_y(s_1, \delta_x^{V})
    \end{equation}
    
    \item[$T_{A2}$:] a back-running transaction exchanges $\delta_y^{A2} = \delta_y^{A1} - c_y(s_2, \delta_y^{A2})$ for $o_x^{A2}$ of $X$ and is planned to execute after $T_V$ (e.g.\ by paying a lower transaction fee than $T_V$). 
    $T_{A2}$ effectively closes the adversary's position that was opened by $T_{A1}$ (cf. Equation~\ref{eq:lt-a2-1} and~\ref{eq:lt-a2-2}).
    \begin{equation}\label{eq:lt-a2-1}
    s_2\xrightarrow{\text{TransactXForY}(\delta_y^{A2})} s_3=(x_3, y_3),\quad \delta_x^{A2} = x_2 - x_3
    \end{equation}
    \begin{equation}\label{eq:lt-a2-2}
    x_3 = \frac{x_2 y_2}{y_3 - c_y(s_2, \delta_y^{A2})} + c_x(s_0, \delta_y^{A2}),\quad y_3 = y_2 + \delta_y^{A2}
    \end{equation}
\end{description}

Transaction $T_{A2}$ swaps the asset $Y$ from transaction $T_{A1}$ in exchange for asset $X$. 
The corresponding profit is determined by comparing the input from $T_{A1}$ and the output $T_{A2}$ of asset $X$ (cf. Equation~\ref{eq:attack1-profit}).

\begin{equation}\label{eq:attack1-profit}
\text{profit} = \delta_x^{A2} - \delta_x^{A1}
\end{equation}

After the attack, the state of adversary becomes $(\delta_x^{A2}, 0)$. Thus, provided the profit exceeds the costs (e.g. transaction fees and equipment costs), a rational adversary would undertake the attack.

\subsection{Liquidity Provider Attacks Taker}\label{app:provider-attacks-taker}
We use $(x_N, y_N)$ to denote the state of an AMM exchange at block $N \in \mathbb{Z}$, where $x_N$, $y_N$ are the amounts of asset $X$ and $Y$ in the liquidity pool (cf.\ Definition~\ref{def:amm-state}). 
The liquidity provider owns a share of $L \in [0,1]$ liquidity if it deposits $\delta_{x, N} = x_N \frac{L}{1 - L}$ of asset $X$ and $\delta_{y, N}= y_N \frac{L}{1 - L}$ of asset $Y$ into the liquidity pool. 
We use $Z_N(x,y)$ to denote the USD value of assets at block $N$, where $x$ is the amount of asset $X$ and $y$ is the amount of asset $Y$. Given two blockchain blocks $m, n$, where the respective AMM states differ, the deposit of $z_m = Z_m(x_m \frac{L}{1 - L}, y_m \frac{L}{1 - L})$ at block $m$, or $z_n = Z_n(x_n \frac{L}{1 - L}, y_n \frac{L}{1 - L})$ at block $n$ both results in a share of $L$ liquidity in the respective market $X$/$Y$. For the following section, we set $z_m > z_n$.

\subsubsection{Attack Profitability}
We proceed by defining profitability for an adversary. $L_x, L_x^*, L_y, L_y^* \in [0, 1[$ denote the proportion of assets $X$ and $Y$ held by the adversary in the liquidity pools before and after the attack. Analogously, $x^A, x^{A*}, y^A, y^{A*} \in \mathbb{Z}^+$ denote the amounts of asset $X$ and $Y$ held by the adversary before and after the attack. 

\begin{equation}\label{eq:attack_a2_state_transition}
    (L_x, L_y, x^A, y^A) \xrightarrow{\text{attack}} (L_x^*, L_y^*, x^{A*}, y^{A*})
\end{equation}

The state transition (cf.\ Equation~\ref{eq:attack_a2_state_transition}) is profitable to the adversary, if the following conditions hold:

\begin{itemize}
    \item $L_x^* \nless L_x,\quad L_y^* \nless L_y,\quad x^{A*} \nless x^A,\quad y^{A*} \nless y^A,\quad$ and
    \item At least one of the following holds:
    
    $L_x^* > L_x,\:$ or $\:L_y^* > L_y,\:$ or $\:x^{A*} > x^A,\:$ or $\:y^{A*} > y^A$
\end{itemize}

Let $x^*$ and $y^*$ denote the amounts of asset $X$ and $Y$ in liquidity pools after the attack. The corresponding profit is determined by comparing adversary's states before and after the attack (cf. Equation \ref{eq:profit_attack_2_x}, \ref{eq:profit_attack_2_y}, \ref{eq:profit-attack-2}).
\begin{equation}\label{eq:profit_attack_2_x}
    \Delta_x^A = x^{A*} - x^A + (L_x^* - L_x) x^*
\end{equation}
\begin{equation}\label{eq:profit_attack_2_y}
    \Delta_y^A = y^{A*} - y^A + (L_y^* - L_y) y^*
\end{equation}
\begin{equation}\label{eq:profit-attack-2}
    \text{profit} = Z(\Delta_x^A, \Delta_y^A)
\end{equation}

\subsubsection{Attack Execution}
We now consider an adversarial liquidity provider that owns a share $L_x x_0, L_y y_0$ of the total liquidity pool of a AMM $X$/$Y$ market. The victim's transaction $T_V$ transacts asset $X$ for $Y$. 
If the adversary does not front-run $T_V$, the AMM state changes according to Equation \ref{eq:lp-v-noattack-1}, \ref{eq:lp-v-noattack-2}. 
In that case, the adversary $A$ receives a commission fee $c$, as stated in Equation \ref{eq:lp-v-noattack-3}, \ref{eq:lp-v-noattack-4}.

\begin{equation}\label{eq:lp-v-noattack-1}
s_0\xrightarrow{\text{TransactXForY}(\delta_x^{V})} s_1=(x_1, y_1),\quad \delta_y^{V} = y_0 - y_1
\end{equation}
\begin{equation}\label{eq:lp-v-noattack-2}
x_1 = x_0 + \delta_x^{V},\quad y_1 = \frac{x_0 y_0}{x_1 - c_x(s_0, \delta_x^{V})} + c_y(s_0, \delta_x^{V})
\end{equation}
\begin{equation}\label{eq:lp-v-noattack-3}
c_x^{A, T_V} = L \times c_x(s_0, \delta_x^{V}),\quad c_y^{A, T_V} = L \times c_y(s_0, \delta_x^{V})
\end{equation}
\begin{equation}\label{eq:lp-v-noattack-4}
(L_x, L_y, x^A, y^A) \xrightarrow{} (L_x, L_y, x^A + c_x^{A, T_V}, y^A + c_y^{A, T_V})
\end{equation}

This liquidity provider attempts to gain a profit through the following order of transactions. 

\begin{description}
    \item[$T_{A1}$:] {a front-running transaction executed before $T_V$ (e.g.\ by paying a higher transaction fee than $T_V$). 
    $T_{A1}$ withdraws $(L_x x_0, L_y y_0)$ from the liquidity pool $(x_0, y_0)$ and results in the state changes of Equation~\ref{eq:lp-a1-1} and~\ref{eq:lp-a1-2}.}
    \begin{equation}\label{eq:lp-a1-1}
    s_0\xrightarrow{\text{RemoveLiquidity}(L_x x_0, L_y y_0)}s_1 = (x_1, y_1)
    \end{equation}
    \begin{equation}\label{eq:lp-a1-2}
    x_1 = x_0 - L_x x_0,\quad y_1 = y_0 - L_y y_0
    \end{equation}
    
    \item[$T_V$:] {a victim's transaction modifies the state according to Equation \ref{eq:lp-v-1} and \ref{eq:lp-v-2}.}
    \begin{equation}\label{eq:lp-v-1}
    s_1\xrightarrow{\text{TransactXForY}(\delta_x^{V})} s_2=(x_2, y_2),\quad \delta_y^{V} = y_1 - y_2
    \end{equation}
    \begin{equation}\label{eq:lp-v-2}
    x_2 = x_1 + \delta_x^{V},\quad y_2 = \frac{x_1 y_1}{x_2 - c_x(s_1, \delta_x^{V})} + c_y(s_1, \delta_x^{V})
    \end{equation}
    
    \item[$T_{A2}$:] {a back-running transaction executes after $T_V$ (e.g.\ by paying a lower transaction fee than $T_V$). 
    $T_{A2}$ adds back liquidity for the adversary to maintain the same proportion of overall liquidity $(L_x, L_y)$, modifying the state according to Equation~\ref{eq:lp-a2-1} and~\ref{eq:lp-a2-2}.}
    \begin{equation}\label{eq:lp-a2-1}
    s_2\xrightarrow{\text{AddLiquidity}(\frac{L_x x_2}{1-L_x}, \frac{L_y y_2}{1-L_y})}s_3 = (x_3, y_3)
    \end{equation}
    \begin{equation}\label{eq:lp-a2-2}
    x_3 = \frac{x_2}{1-L_x},\quad y_3 = \frac{y_2}{1-L_y}
    \end{equation}
    
    \item[$T_{A3}$:] executed after $T_{A2}$, $T_{A3}$ rebalances the AMM assets by converting $Y$ to $X$, such that the adversary retains the same amount of asset $X$ as before the attack (i.e.\ the adversary holds $x^A + c_x^{A, T_V}$ after the attack). 
    This rebalancing process is necessary because the amount of asset $X$ added to liquidity pool in $T_{A2}$ exceeds the amount withdrawn from $T_{A1}$. 
    $T_{A3}$ modifies the state according to the Equations \ref{eq:lp-a3-1}, \ref{eq:lp-a3-2} and \ref{eq:lp-a3-3}.
    
    \begin{equation}\label{eq:lp-a3-1}
    s_3\xrightarrow{\text{TransactXForY}(\delta_y^{A3})} s_4=(x_4, y_4),\quad \delta_x^{A3} = x_3 - x_4
    \end{equation}
    \begin{equation}\label{eq:lp-a3-2}
    x_4 = \frac{x_3 y_3}{y_4 - c_y(s_3, \delta_y^{A})} + c_x(s_3, \delta_y^{A}),\quad y_4 = y_3 + \delta_y^{A3}
    \end{equation}
    {
    \begin{equation}\label{eq:lp-a3-3}
    \delta_x^{A3} = \frac{L_x x_2}{1-L_x} + c_x^{A, T_V} - L_x x_0
    \end{equation}
    }
\end{description}

The overall state change of this attack is described in Equation \ref{eq:attack_a2_state_transition_2}. 
At the end of this attack, the adversary $A$ has managed to retain $L$ proportion of the total liquidity as a result of $T_{A2}$, retains the same total amount of asset $X$ (as a result of $T_{A3}$) but increases their holding of asset $Y$, generating a profit. 
Equation~\ref{eq:attack_a2_profit} yields the profit.

\begin{equation}\label{eq:attack_a2_state_transition_2}
    (L_x, L_y, x^A, y^A) \xrightarrow{\text{attack}} (L_x, L_y, x^A + c_x^{A, T_V}, y^{A*})
\end{equation}

\begin{equation}\label{eq:attack_a2_profit}
\text{profit} = y^{A*} - (y^A + c_y^{A, T_V})
\end{equation}

\begin{figure}[htb!]
    \begin{center}
    \includegraphics[width = 0.95\columnwidth]{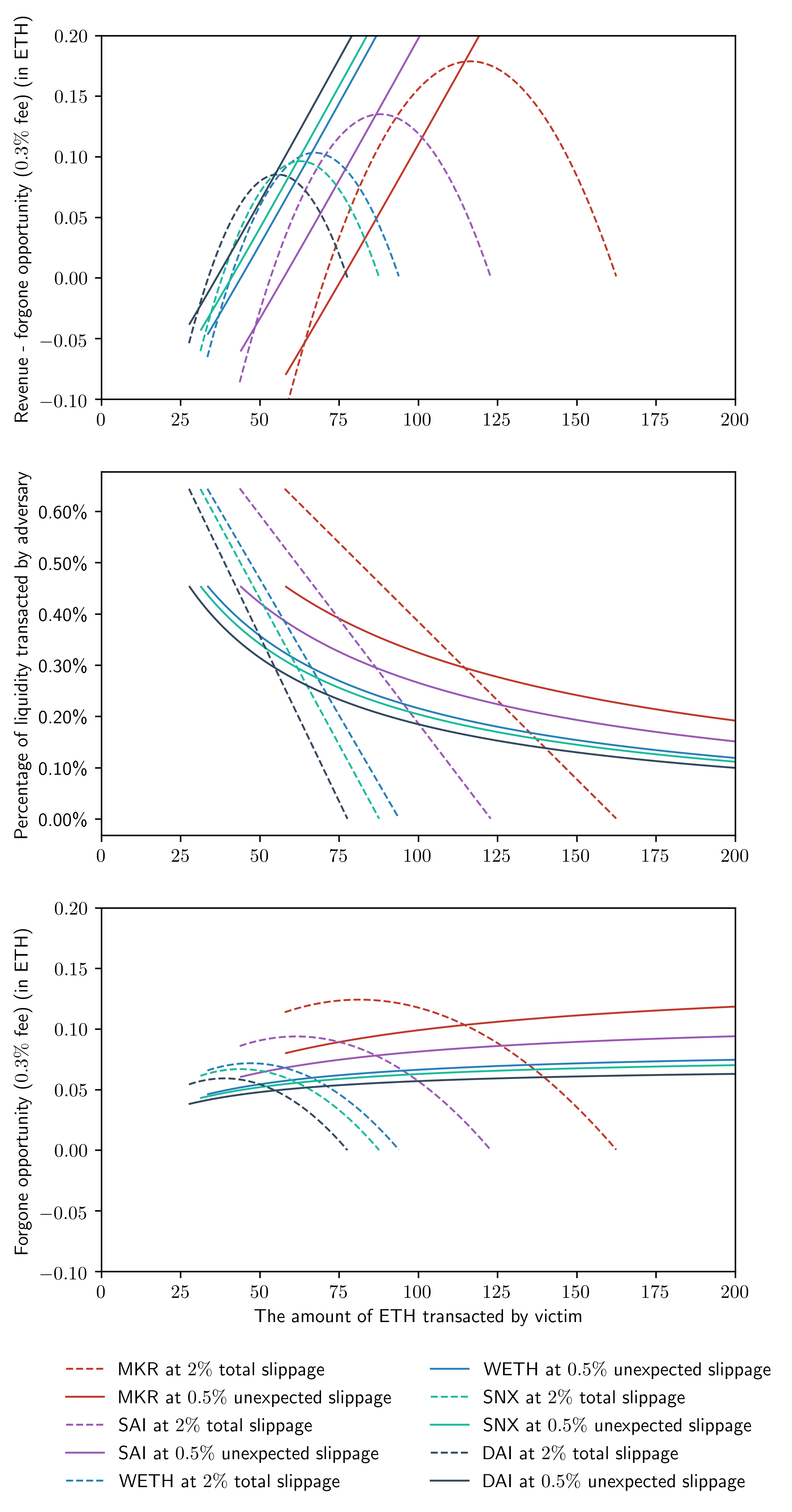}
    \end{center}
    \caption{Optimal adversarial revenue under a sandwich attack by a liquidity provider, when $V$ sells assets for ETH on five Uniswap exchanges ($0.3\%$ fee, adversarial break-even at $0.01\ \text{ETH}$). 
    }
    \label{fig:optimal-attack2}
\end{figure}

\begin{figure*}
\centering
\subfigure[Two liquidity provider adversaries]{\label{fig:provider_2_adversaries}\includegraphics[width=0.32\textwidth]{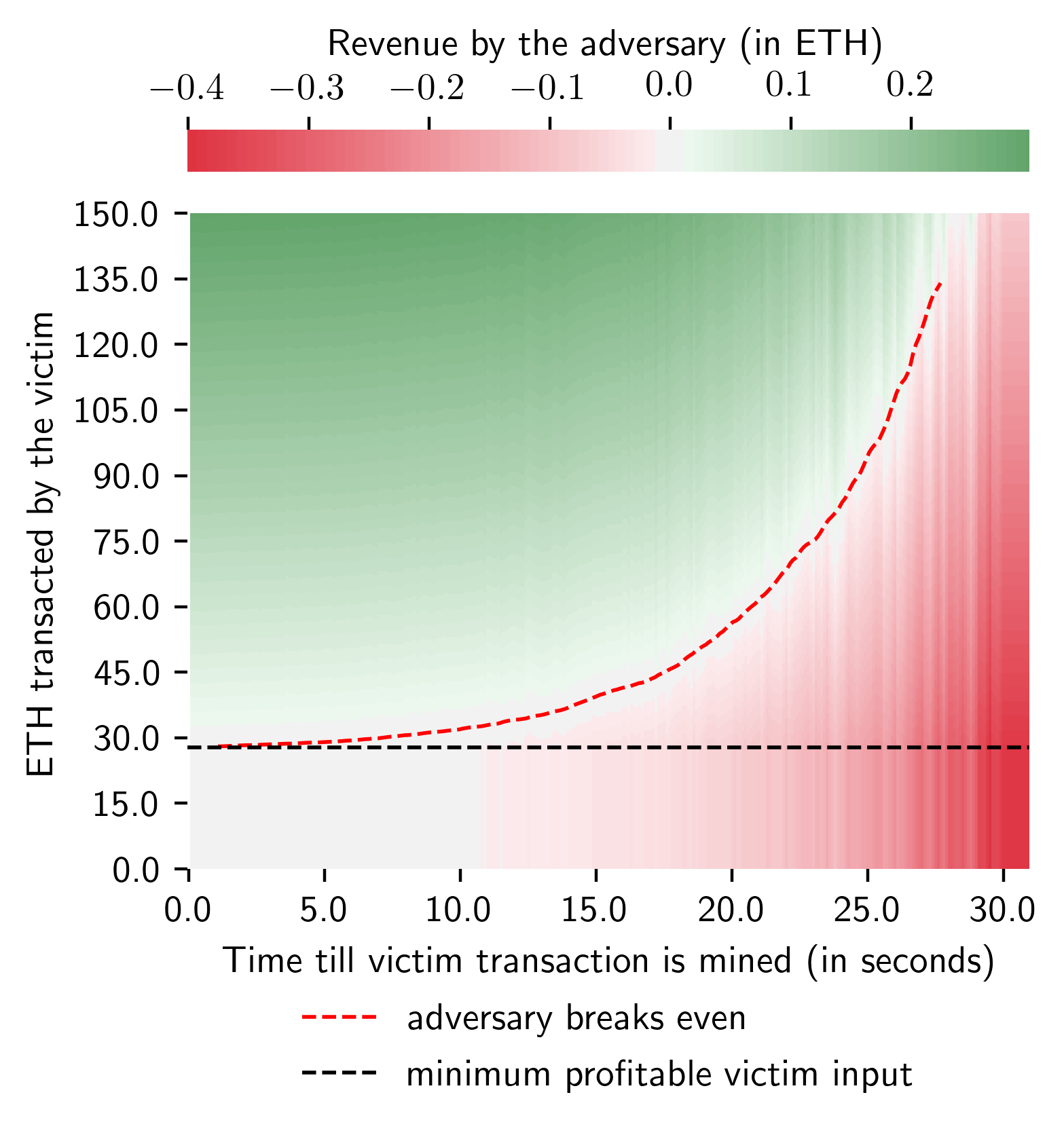}}
\subfigure[Five liquidity provider adversaries]{\label{fig:provider_5_adversaries}\includegraphics[width=0.32\textwidth]{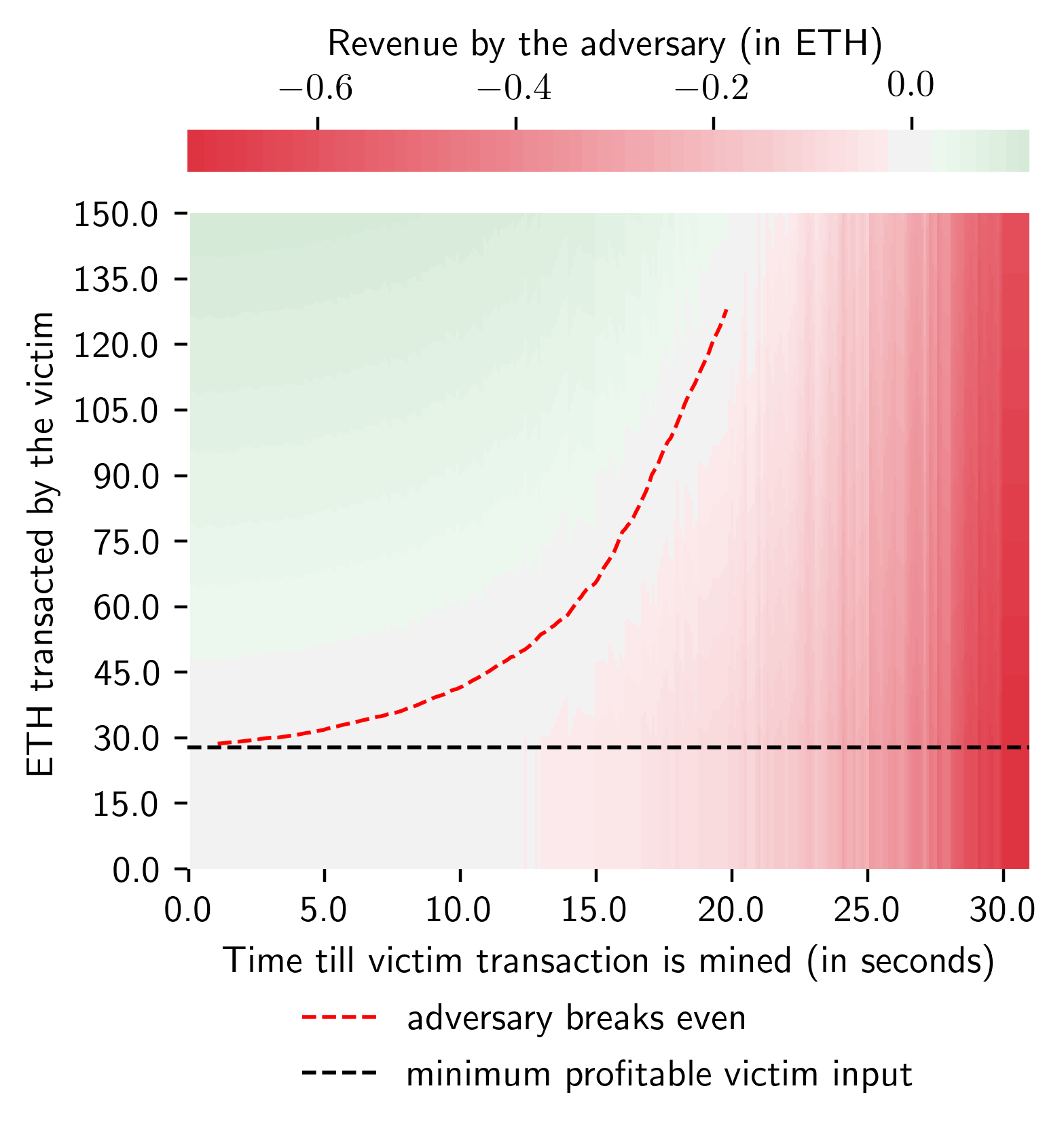}}
\subfigure[Ten liquidity provider adversaries]{\label{fig:provider_10_adversaries}\includegraphics[width=0.32\textwidth]{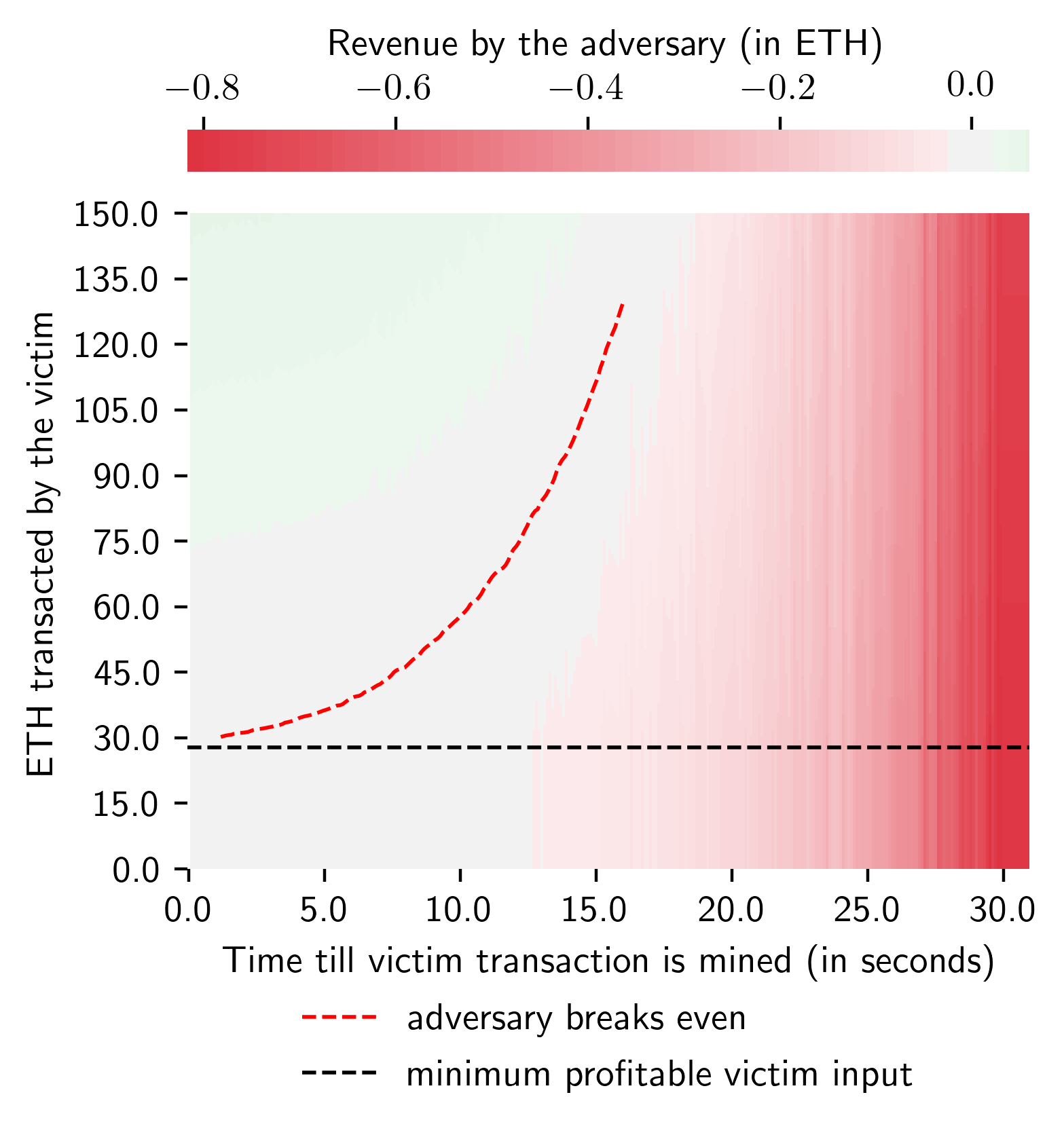}}
\caption{Simulated sandwich attack by $2$, $5$ and $10$ competing adversarial liquidity provider on a taker (Uniswap, block $9$M, $0.3\%$ fees, $0.5\%$ unexpected slippage).}
\label{fig:provider_multiple_adversary}
\end{figure*}

\section{Additional Details For Liquidity Provider}

Figure~\ref{fig:optimal-attack2} quantifies the optimal adversarial revenue by a liquidity provider, given no other attacks are running.

Figure~\ref{fig:provider_multiple_adversary} shows the expected profit of an adversarial liquidity provider (cf. Section~\ref{sec:analytical-evaluation}) given 2, 5, and 10 adversaries, on the Uniswap DAI market at block 9M. The victim transaction is fixed with a $0.5\%$ slippage protection. The minimum profitable victim input with a single adversary is $27.8$ ETH ($4,141.37$ USD).

\section{Possible Mitigations}\label{sec:defences}
In this section we elaborate on how users could protect their trades from sandwich attacks, and we discuss possible AMM design changes to strengthen their resilience.

\subsection{AMM Specific Settings}
We identify two primary protection possibilities that could be adopted given the exsting AMM DEX designs (such as Uniswap).

\subsubsection{Allow for Zero Slippage}
The simplest protection for a user, is to not allow its trades to pay any slippage. If an adversary were to front-run, the user's trade execution will fail. The user would not overpay the trade due to its slippage protection, but the user still is liable to pay blockchain transaction fees. This protection solution moreover is not scalable in terms of trades per second, as it would limit the number of trades an AMM DEX can execute to $1$ transaction per blockchain block.

\subsubsection{Limit Liquidity Taker's Input}
Recall that sandwich attacks are only profitable if the victim's input amount is above a threshold, which we named it as the minimum profitable victim input (cf.\ Section~\ref{sec:analytical-evaluation}). This threshold depends on both the total liquidity of the market and the AMM DEX's design and configuration (pricing formula, fees and etc.). One possible protection is to disable transactions above the minimum profitable victim input in the smart contracts.

\subsection{Cryptography-based Defences}
In the following we discuss possible cryptography based defence techniques against sandwich attacks.

\subsubsection{Multi-Party Computation (MPC)}
The sandwich attack is possible because the current consensus protocol used in Ethereum fails to protect the actual ordering of the transactions from adversarial manipulations (i.e. fee manipulations). Thus, it is tempting for one to design an AMM DEX that has a authorized set of ``trusted'' nodes to faithfully sequence actions from liquidity takers before getting the actions executed by the smart contract. This proposal, however, makes AMM DEX like Uniswap no longer permissionless, and the authorized set can again manipulate the order of the transactions. To address the later, Kelkar~\emph{et al.}~\cite{mahimna-2020-order-fairness-BFT} propose a new set of Byzantine Consensus protocols, that achieve a fair ordering of received transactions. Thus, one can require the authorized set of nodes to run such protocol to achieve order fairness among transactions.  

\subsubsection{Commit-and-Reveal Protocols}
A commitment scheme is a two-round protocol that allows one to commit to chosen values (i.e. function, input) while keeping those values hidden from others (\textit{hiding}) during the committing round, and later during the revealing round, s/he can decide to reveal the committed value. The commitment schemes are \textit{binding} if and only if the party cannot change the value after committing to it. We briefly discuss the use of commitment scheme to prevent front-running in the following.

\paragraph{Standard Commit-and-Reveal Protocol}
    To prevent the sandwich attack in AMM DEX, one can use commitment schemes to sequence actions of traders during committing round and execute actions during revealing round. In particular, during committing round, traders commit to function calls (i.e. \TransactXY, \addliquidity, \removeliquidity) via commitment transactions, and the ordering of function calls is determined based on the order of commitments while the function calls are hidden due to the \textit{hiding} property of a commitment scheme. In the revealing round, parties can decide to reveal the function calls, and the AMM DEX will execute the transactions according to the order of the commitments appeared in the committing round. 

    One of the limitations of commit-and-reveal protocol is its usability, as it requires participants to be aware of both rounds of the protocol to complete their actions. Another limitation is that adversary is still able to \textit{probabilistically} perform sandwich attack. In particular, because the committing round is transparent, the adversary can see other traders' commitments and commits several transactions before and after the commitments of honest traders. S/he can reveal only those transactions that are profitable. 

\paragraph{Commit-and-Reveal Protocol tailored for the Ethereum Blockchain}
    To hide the committing phase from the adversary, Breidenbach~\emph{et al.}~\cite{breindenbach2018enter} proposed submarine commitments via contract creation in Ethereum (i.e. $\texttt{CREATE2}$ EVM opcode~\cite{create2-eip}). The essence of this approach is to hide commitment transaction among newly generated Ethereum addresses. In particular, a submarine commitment scheme contains the following phases:
    
\textbf{Committing Phase:} to \textit{commit} in a submarine commitment scheme, the liquidity taker with address $\mathsf{Addr_{Taker}}$ posts a transaction $T_\mathsf{com}$ that sends some fund, $\texttt{val}$, to an address $\mathsf{Addr_{com}}$. $\mathsf{Addr_{com}}$ is a commitment of the form: 
\begin{equation*}
    \mathsf{Addr_{com}}= H(\mathsf{Addr_{AMM\text{-}DEX}},H(\mathsf{Addr_{Taker}}, \mathsf{key}), code)
\end{equation*}
where $H(\cdot)$ is Keccak-256, $\mathsf{key}$ is the transaction specific key to AMM DEX (e.g.\ a concatenation of action $\transactXY$, input $\delta_x$, and 256-bit randomness $r$), and $code$ is the EVM init code of the \textit{refund} contract that can send any money received to the $\mathsf{Addr_{AMM\text{-}DEX}}$. 

\textbf{Revealing Phase:} To reveal, $\mathsf{Addr_{Taker}}$ sends to $\mathsf{Addr_{AMM\text{-}DEX}}$, key value $\mathsf{key}=({\color{blue}{\mathsf{action}}}||input||r)$, the transaction data $T_\mathsf{com}$, the $\mathsf{commitBlock}$ (the block number includes $T_\mathsf{com}$), and a Merkle-Partricia proof, $\pi_{T_\mathsf{com}}$ which proves the membership of $T_\mathsf{com}$ in $\mathsf{commitBlock}$. With $\mathsf{commitBlock}$, $\pi_{T_\mathsf{com}}$, and $T_\mathsf{com}$, $\mathsf{Addr_{AMM\text{-}DEX}}$ verifies that $T_\mathsf{com}$ occurred in $\mathsf{commitBlock}$. And after learning $\mathsf{key}$, $\mathsf{Addr_{AMM\text{-}DEX}}$ can recompute $\mathsf{\mathsf{Addr_{com}}}$, verifies the deposit balance $\texttt{val}$, and proceeds with $\mathsf{action}$ and $input$ (i.e. $\transactXY(\delta_x)$). 

\textbf{Deposit Collection Phase:} from $code$, $\mathsf{Addr_{AMM\text{-}DEX}}$ can use $\texttt{CREATE2}$ opcode to create an instance of the \textit{refund} contract at $\mathsf{Addr_{com}}$ and collect $\$\texttt{val}$ from $\mathsf{Addr_{com}}$.

\subsubsection{Confidential Transactions}
Another potential attempt to mitigate front-running attack is to hide the details of the transaction sent to AMM DEX by adapting several techniques~\cite{sasson2014zerocash,zether-bunz-2020,zexe-2020-sp,asiacrypt-2019-quisquis} for confidential transactions. 
However, as pointed out by Eskandari et al.~\cite{eskandari-sok-frontrun-fc2019}, to prevent front-running, one needs to hide:
\begin{itemize}
    \item[(1)] The name of the functions (i.e. \TransactXY, \addliquidity, \removeliquidity) being invoked 
    \item[(2)] The parameters supplied to the functions (i.e. $\delta_x, \delta_y$)
    \item[(3)] The current state of the DEX (i.e. $(x,y)$).
\end{itemize} 
While systems like Hawk~\cite{kosbahawk} and Ekiden~\cite{cheng2019ekiden} try to achieve all three properties for arbitrary functions, they rely on off-chain components (i.e.\ trusted execution environment) for maintaining encrypted states and proving the correctness of state transitions. On the other hand, a proposal for Ethereum blockchain, Zether~\cite{zether-bunz-2020}, tries to achieve $(2,3)$ for a specific type of function (i.e.\ money transferring). The states in Zether are the ElGamal encryptions of account's balances. The state transitions can be made due to the correctness of the non-interactive zero-knowledge proof (NIZK) system used in their construction and the (additively) homomorphic properties of Elgamal encryption.

However, recall that in an AMM DEX, an action like $\TransactXY(\delta_x)$ requires the contract to send an $\delta_y$ of asset $Y$ back to the liquidity taker where $\delta_y$ is computed based on the pricing function $f(\cdot)$ and the current pool's state $(x,y)$. Therefore, if one decides to use a system that relies on NIZK systems like Zether for AMM DEX with hidden state, the trusted off-chain components are needed to generate cryptographic proofs from the unencrypted state to trigger the transferring back function.

Moreover, regardless of the privacy techniques used for hiding transaction details and pool's states, we show that it is impossible to achieve the confidentiality for the third property for a constant product AMM. In particular, at the initial pool state $s=(x,y)$, an adversary can issue two consecutive actions, $\TransactXY(\delta_x),\TransactXY(\delta_x')$, and it obtains $\delta_y$ and $\delta'_y$ determined by the current state $(x,y)$.
The adversary can solve the system of two equations and two unknown $(x,y)$ to determine the current pool state.



\clearpage

\end{document}